\documentclass[a4paper,10pt]{article}
\usepackage[T1]{fontenc}
\usepackage[utf8]{inputenc}
\usepackage{mathpazo}
\usepackage[english]{babel}
\usepackage{amsmath, amsfonts, amsthm}

\usepackage{geometry}
\usepackage[font=small,labelfont=bf]{caption}
\usepackage{pgf}
\usepackage{tikz}
\usetikzlibrary{calc, shapes, shapes.geometric, shapes.symbols}
\usepackage{csquotes}
\usepackage{caption}
\usepackage[aboveskip=-0.5em]{subcaption}
\usepackage{authblk}
\usepackage{enumitem}
\usepackage[backend=bibtex,doi=true,style=numeric,maxnames=6]{biblatex}
\usetikzlibrary{external}
\newcommand\mysubref[1]{\textbf{\subref{#1}:}}

\theoremstyle{definition}
\newtheorem{definition}{Definition}

\theoremstyle{plain}
\newtheorem{theorem}[definition]{Theorem}

\renewcommand{\eqref}[1]{equation~(\ref{eq:#1})}

\newcommand{\Figref}[1]{Figure~\ref{fig:#1}}
\newcommand{\figref}[1]{Figure~\ref{fig:#1}}
\newcommand{\ie}{i.e.\ }
\newcommand{\eg}{e.g.\ }
\newcommand{\Eg}{E.g.\ }
\newcommand{\etal}{et al.\ }
\newcommand{\secref}[1]{Section~\ref{sec:#1}}
\newcommand{\Secref}[1]{Section~\ref{sec:#1}}
\newcommand{\thmref}[1]{Theorem~\ref{thm:#1}}

\newcommand{\defref}[1]{Definition~\ref{def:#1}}

\newcommand{\defword}[1]{\emph{#1}}
\newcommand{\ZZ}{\ensuremath{\mathbb{Z}}}
\newcommand{\RR}{\ensuremath{\mathbb{R}}}
\newcommand{\NN}{\ensuremath{\mathbb{N}}}
\newcommand{\suchthat}{\ensuremath{\, \mid \,}}
\newcommand{\given}{\ensuremath{\, \mid \,}}
\newcommand{\iso}{\ensuremath{\cong}}

\newcommand{\inv}{\ensuremath{^{-1}}}
\newcommand{\Cf}{C.f.\ }
\newcommand{\rmd}[0]{\ensuremath{\;\mathrm{d}}}
\newcommand{\id}[0]{\ensuremath{\mathrm{id}}}

\newcommand{\Cech}{Čech}
\newcommand{\Ccplx}{\ensuremath{\check{C}}}

\newcommand{\rat}[1]{\includegraphics[width=0.5cm,angle=#1]{figures/mouse.pdf}}

\newcommand{\localfigname}[0]{NONE}
\newcommand{\localfigprefix}[0]{NONE}
\DeclareMathOperator{\VR}{VR}

\DeclareMathOperator{\corr}{corr}
\DeclareMathOperator{\corravg}{\overline{corr}}
\DeclareMathOperator{\flag}{fl}
\DeclareMathOperator{\OC}{OC}
\DeclareMathOperator{\shuf}{sh}
\DeclareMathOperator{\pr}{pr}

\tikzset{parallelogram/.style={trapezium, trapezium left angle=70, trapezium right angle=-70}}
\tikzset{twosimp/.style={fill opacity=0.6,fill=gray,draw opacity=0.9}}
\tikzset{threesimp/.style={fill opacity=0.8,fill=blue!60,draw opacity=0.9}}

\newcommand{\overbar}[1]{\mkern 1.5mu\overline{\mkern-1.5mu#1\mkern-1.5mu}\mkern 1.5mu}
\usepackage{bm}
\newcommand{\boldchi}{\ensuremath{\bm{\chi}}}
\newcommand{\fastfig}[2]{
  \includegraphics{figures/#1/arxiv/#2.pdf}
}

\title{Using persistent homology to reveal hidden information in neural data}
\author[1]{Gard Spreemann}
\author[2]{Benjamin Dunn}
\author[1]{Magnus Bakke Botnan}
\author[1]{Nils~A.~Baas}
\affil[1]{\{gard.spreemann, magnus.botnan, nils.baas\}@math.ntnu.no\\ Department of Mathematical Sciences, Norwegian University of Science and Technology, 7491 Trondheim, Norway}
\affil[2]{benjamin.dunn@ntnu.no\\ Kavli Institute for Systems Neuroscience, Norwegian University of Science and Technology, 7491 Trondheim, Norway}
\date{\today}

\begin{document}
\maketitle
%\setcounter{tocdepth}{2}
%\tableofcontents

\begin{abstract}
  We propose a method, based on persistent homology, to uncover
  topological properties of a priori unknown covariates of neuron
  activity. Our input data consist of spike train measurements of a
  set of neurons of interest, a candidate list of the known stimuli
  that govern neuron activity, and the corresponding state of the
  animal throughout the experiment performed. Using a generalized
  linear model for neuron activity and simple assumptions on the
  effects of the external stimuli, we infer away any contribution to
  the observed spike trains by the candidate stimuli. Persistent
  homology then reveals useful information about any further, unknown,
  covariates.
\end{abstract}

\newpage

\section{Introduction} \label{sec:intro}

\begin{figure}[htbp]
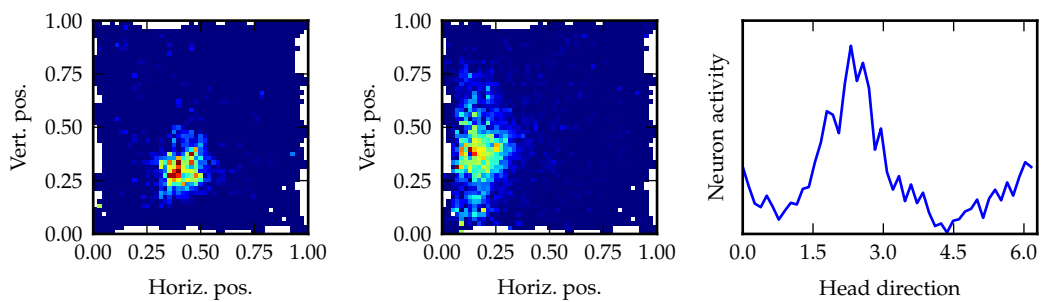

  \centering
  \fastfig{place-fields}{placefield-9_50}\hspace{-1em}
  \fastfig{place-fields}{placefield-30_50}\hspace{-1em}
  \fastfig{place-fields}{placefield-12-head_50}
  \caption{The firing activity of three neurons as a rat explores a
    square box. For the two leftmost neurons the activity is plotted
    against spatial position (both axes) while the activity for the
    rightmost one is plotted against head direction (horizontal
    axis). The regions of elevated activity, the spatial and ``head
    direction state space'' place fields, are clearly visible. In the
    spatial plots, the cells were least active in the blue areas, and
    most in the red. White areas were not visited by the rat during
    the experiment. Activity and position data for the plots come from
    \cite{buzsaki2013data, mizuseki2009theta}.} \label{fig:placefield}
\end{figure}

Due to its apparent simplicity, physical space has long served as a
model system for internally generated representations in the
brain~\cite{tolman1948cognitive}. In their pioneering
work~\cite{okeefe1971}, O'Keefe and Dostrovsky discovered in the
hippocampi of rats certain neurons, called \emph{place cells}, that
seemed to be active at a level far above their baseline when the
animal was located in a specific region in space. These regions of
elevated activity are known as \emph{place fields}. It has also been
demonstrated~\cite{muller1996review, save2000sensory, rubin2014head}
that neurons tune not only to spatial position, but also to other
external covariates, such as for example head direction. Place fields
thus do not exist only for an animal's physical surroundings --- we
shall rather think of them as regions in an abstract state space, thus
generalizing to the more fundamental concept of neural coding. An
example of the place fields of three cells recorded in an experiment
by Buzsáki \etal{}~\cite{buzsaki2013data, mizuseki2009theta} can be
seen in \figref{placefield}.

If a collection of neurons divide the animal's state space into
reasonably nice place fields, it is perhaps not surprising that the
activity of these neurons somehow encode information about the state
space. A question to pose is then how much of that space can be
recovered from just recordings of the electrical activity of the
neurons. We will consider this question from a novel point of view in
which the state space is partially known, and see what properties of
the unknown parts can be glanced from such recordings. This, in turn,
will give us information about unknown external covariates of neuron
activity by using a generalized linear model
(GLM)~\cite{mccullagh1989generalized} to infer away the contributions
of any known covariates. GLMs have recently been applied in
neuroscience as a structured method to uncover dependencies as well as
structure in data~\cite{pillow2008spatio, truccolo2005point,
  roudi2015multi}.

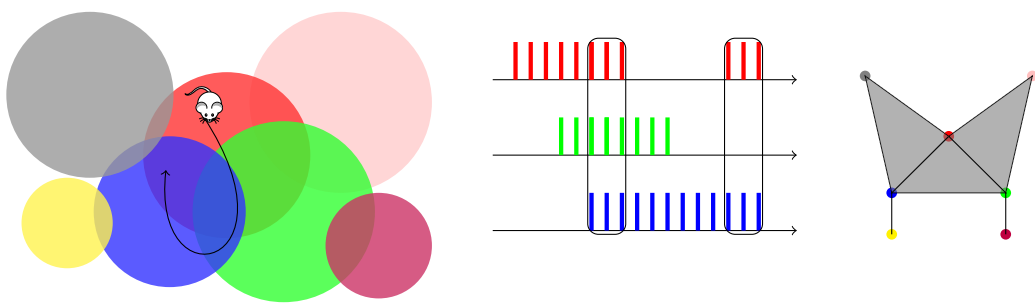
\begin{figure}[htbp]
  \centering
  \begin{tikzpicture}
    \begin{scope}
      \fill[color=pink!80, fill opacity=0.8] (4.5, -1.3) circle (1.2);
      \fill[color=red!80,fill opacity=0.8] (3, -2) circle (1.1);
      \fill[color=green!80,fill opacity=0.8] (3.75, -2.75) circle (1.2);
      \fill[color=blue!80,fill opacity=0.8] (2.25, -2.75) circle (1.0);
      \fill[color=yellow!80,fill opacity=0.8] (0.9, -2.9) circle (0.6);          
      \fill[color=gray!80, fill opacity=0.8] (1.2, -1.2) circle (1.1);
      \fill[color=purple!80, fill opacity=0.8] (5, -3.2) circle (0.7);
      \draw[->] (2.7, -1.5) .. controls (4, -3.5) and (2, -4) ..  (2.2, -2.2);
      \node[anchor=mid] () at (2.7, -1.5) {\rat{0}};      
    \end{scope}
    \begin{scope}[xshift=6.5cm, yshift=-1cm]
      \draw[->] (0,0) -- (4, 0);
      \draw[->] (0,-1) -- (4,-1);
      \draw[->] (0,-2) -- (4, -2);
      
      \draw[ultra thick,red]   (0.3, 0) -- +(0, 0.5);
      \draw[ultra thick,red]   (0.5, 0) -- +(0, 0.5);
      \draw[ultra thick,red]   (0.7, 0) -- +(0, 0.5);
      
      \draw[ultra thick,red]   (0.9, 0) -- +(0, 0.5);
      \draw[ultra thick,red]   (1.1, 0) -- +(0, 0.5);
      \draw[ultra thick,red]   (1.3, 0) -- +(0, 0.5);
      \draw[ultra thick,red]   (1.5, 0) -- +(0, 0.5);
          
      \draw[ultra thick,green]   (0.9, -1) -- +(0, 0.5);
      \draw[ultra thick,green]   (1.1, -1) -- +(0, 0.5);
      \draw[ultra thick,green]   (1.3, -1) -- +(0, 0.5);
      \draw[ultra thick,green]   (1.5, -1) -- +(0, 0.5);
      
      \draw[ultra thick,red]   (1.3, 0) -- +(0, 0.5);
      \draw[ultra thick,red]   (1.5, 0) -- +(0, 0.5);
      \draw[ultra thick,red]   (1.7, 0) -- +(0, 0.5);
      \draw[ultra thick,green]   (1.3, -1) -- +(0, 0.5);
      \draw[ultra thick,green]   (1.5, -1) -- +(0, 0.5);
      \draw[ultra thick,green]   (1.7, -1) -- +(0, 0.5);
      \draw[ultra thick,blue]   (1.3, -2) -- +(0, 0.5);
      \draw[ultra thick,blue]   (1.5, -2) -- +(0, 0.5);
      \draw[ultra thick,blue]   (1.7, -2) -- +(0, 0.5);
      
      \draw[ultra thick,green]   (1.9, -1) -- +(0, 0.5);
      \draw[ultra thick,green]   (2.1, -1) -- +(0, 0.5);
      \draw[ultra thick,green]   (2.3, -1) -- +(0, 0.5);
      \draw[ultra thick,blue]   (1.9, -2) -- +(0, 0.5);
      \draw[ultra thick,blue]   (2.1, -2) -- +(0, 0.5);
      \draw[ultra thick,blue]   (2.3, -2) -- +(0, 0.5);

      \draw[ultra thick,blue]   (2.5, -2) -- +(0, 0.5);
      \draw[ultra thick,blue]   (2.7, -2) -- +(0, 0.5);
      \draw[ultra thick,blue]   (2.9, -2) -- +(0, 0.5);

      \draw[ultra thick,red]   (3.1, 0) -- +(0, 0.5);
      \draw[ultra thick,red]   (3.3, 0) -- +(0, 0.5);
      \draw[ultra thick,red]   (3.5, 0) -- +(0, 0.5);
      \draw[ultra thick,blue]   (3.1, -2) -- +(0, 0.5);
      \draw[ultra thick,blue]   (3.3, -2) -- +(0, 0.5);
      \draw[ultra thick,blue]   (3.5, -2) -- +(0, 0.5);
      
      \draw[rounded corners] (1.25, 0.55) -- (1.75, 0.55) -- (1.75, -2.05) -- (1.25, -2.05) -- cycle;
      \draw[rounded corners] (3.05, 0.55) -- (3.55, 0.55) -- (3.55, -2.05) -- (3.05, -2.05) -- cycle;
    \end{scope}
    \begin{scope}[xshift=9.5cm, yshift=0.25cm]
      \coordinate (a) at (3, -2);
      \coordinate (b) at (3.75, -2.75);
      \coordinate (c) at (2.25, -2.75);
      \coordinate (d) at (2.25, -3.3);
      \coordinate (e) at (1.9, -1.2);
      \coordinate (f) at (3.75, -3.3);
      \coordinate (g) at (4.1, -1.2);
      
      \node[circle, color=red, fill=red, minimum size=4pt, inner sep=0pt] () at (a) {};
      \node[circle, color=green, fill=green, minimum size=4pt, inner sep=0pt] () at (b) {};
      \node[circle, color=blue, fill=blue, minimum size=4pt, inner sep=0pt] () at (c) {};
      \node[circle, color=yellow, fill=yellow, minimum size=4pt, inner sep=0pt] () at (d) {};
      \node[circle, color=gray, fill=gray, minimum size=4pt, inner sep=0pt] () at (e) {};
      \node[circle, color=purple, fill=purple, minimum size=4pt, inner sep=0pt] () at (f) {};
      \node[circle, color=pink, fill=pink, minimum size=4pt, inner sep=0pt] () at (g) {};
      
      \draw[twosimp] (a) -- (b) -- (c) -- cycle;
      \draw[twosimp] (a) -- (c) -- (e) -- cycle;
      \draw[twosimp] (a) -- (g) -- (b) -- cycle;
      \draw (c) -- (d);
      \draw (b) -- (f);
    \end{scope}
  \end{tikzpicture}
  \caption{Place cell cofiring as a proxy for place field
    intersections. As the animal moves along the indicated path
    (\textbf{left}), we might observe the (highly idealized) firing
    events (\textbf{center}) of the corresponding place cells. The
    firing events in the leftmost box are indicative of the triple
    intersection of the red, green and blue place fields, and those in
    the rightmost box are indiciative of the double intersection of
    the red and blue fields. After the space has been thoroughly
    explored, the intersection data obtained from the cofiring of the
    spike trains, define a simplicial complex (\textbf{right}) by
    means of the nerve construction. The simplicial complex and the
    space covered by place fields have the same homotopy type (in this
    example, that of a point).  } \label{fig:cofire-intersections}
\end{figure}

It should be pointed out that in order to compute or visualize the
place fields themselves, as done for \figref{placefield}, the activity
data for the cells must be supplemented by positional and other state
space data, \eg head direction or other external covariates, for the
animal. Without prior knowledge of the nature of the covariates, \ie
prior knowledge of the unknown state space, it is of course not
apparent what data should be recorded. Since it is therefore too much
to ask to know the actual place fields themselves, we take a cue from
topology and instead try for only information on how the place fields
\emph{fit together}. Indeed, if a set of place fields (which are
themselves unknown) intersect, we should be more likely to observe the
corresponding cells firing simultaneously, or nearly so, in
time. \Figref{cofire-intersections} illustrates the idea. This will be
our guide when we build a combinatorial topological space from neuron
activity data in the form of firing event time series (so-called
\defword{spike trains}). We will, in other words, transform our
original time series data --- the collection of spike trains --- into
geometric data, which we will then study with tools from topology.

Cofiring is thus a proxy that gives us approximate knowledge of how
all the place fields intersect (doubly, triply, etc.). We turn to
topology to ask what we can learn from these data. The natural next
step, when armed with intersection data, is to construct the
\emph{nerve} of the place fields. The unfamiliar reader will find the
formal definition in \secref{background}, but intuitively the process
goes as follows: for every cell, draw a point; whenever two place
fields intersect, connect the corresponding two points with a line
segment; whenever three place fields intersect, draw a filled triangle
between the three corresponding points; whenever four place fields
intersect, draw a filled tetrahedron between the four corresponding
points; and so forth. The abstract combinatorial construction so
obtained is one example of a simplicial complex, a kind of topological
space. It is a famous theorem (\thmref{nerve}) in algebraic topology
that this simplicial complex shares an important property with the
space covered by the objects whose intersections we consider, \ie the
state space covered by place fields in our case.

The property that the nerve of the place fields shares (under some
assumptions) with the space covered by the place fields is that of
\emph{homotopy type}. A reader unfamiliar with elementary topics in
algebraic topology may want to see~\cite{hatcher}, or can think of
spaces with the same homotopy type as those that can be continuously
deformed into one another \emph{without tearing}. Thus, a square, a
disk and a point are of the same homotopy type, while they are of
different types from an annulus, which is again of different type from
a sphere. This answers the question of what we may hope to recover
from the cells' firing information: not the full state space (\ie not
the environment's full geometry in a setting with only spatial place
cells), but rather its homotopy type, which again says something about
the kind of \emph{holes} the space has. We should for example, in the
purely spatial case, at the very least be able to tell whether there
is an obstacle, such as an impassable column, somewhere in a box the
rat explores (making it, in the eyes of homotopy, an annulus rather
than a box). Moreover, the circular nature of a covariate such as head
direction also influences the homotopy type of the state space, and is
therefore detectable.

We have so far glossed over what it means for cells to cofire. Clearly
a single cofiring event is not sufficient to declare place fields as
overlapping, as cells may fire spuriously with the animal outside of
place fields. Too strict a notion of cofiring is obviously not good
either. Moreover, looking to the real-world situation in
\figref{placefield}, the \emph{degree} of intersection we should
demand to declare two fields as intersecting is not as clear-cut as
presented in the idealized setting of
\figref{cofire-intersections}. \emph{Persistent homology}, which we
recap in \secref{background}, offers a way to consider all degrees of
cofiring and intersections simultaneously and as one mathematical
object. While not as sharp an invariant as homotopy type, (persistent)
homology still captures the number of holes in the space under
consideration, their dimensionality, and to some extent also their
size.

In summary, the correlations of spike trains allow us to build
combinatorial spaces that reflect topological properties of some
partially unknown animal state space. In the purely spatial setting,
several papers~\cite{curto2008cell, dabaghian2012, giusti2015} have
already demonstrated the feasibility of recovering properties of the
animal's physical environment in this way. We now propose a new method
wherein topological properties of a partially unknown state space are
uncovered by successively accounting for known covariates using a GLM.

\subsection{Contributions} \label{sec:contrib}
Already in~\cite{okeefe1971} it was clear that spatial position is not
the only influence on the firing of certain neurons. Other covariates,
confirmed or suspected, include~\cite{muller1996review,
  save2000sensory, rubin2014head}: head direction, theta wave phase,
behavioral tasks (eating, biting, sleeping, grooming, etc.) and
tactile or olfactory sensory stimuli. In addition to these external
influences, neurons are connected to a set of neighboring neurons, and
are excited or inhibited by the activity of those neighbors.

In this paper, we consider a general setting of neural data governed
by a list of suspected possible covariates (also referred to as
stimuli or influences). We further assume that the researcher supplies
spike train recordings of the relevant neurons, and, in addition,
measurements of the suspected physical covariates throughout the
experiment. We then set out to answer the following questions:
\begin{description}
\item[Q0:] What is the (persistent) homology encoded by the
  covariates, \ie what is the (persistent) homology of the
  corresponding state space?
\item[Q1:] Does the list of suspected covariates adequately explain the
  observations?
\item[Q2:] If not, do the covariates missing from the list of
  candidates encode non-trivial homological information?
\end{description}

The purely spatial version of question~Q0 has already been examined by
others~\cite{curto2008cell, dabaghian2012, giusti2015}. Standard
spectral methods, for example using Wigner's semicircle
law~\cite{wigner1958} on the correlations of neurons, may be
sufficient to answer question~Q1. Still, we will in this paper
approach that question using persistent homology, as done
in~\cite{giusti2015}. We believe that question~Q2 has not been
considered before, and that its answers can be useful to
neuroscientists. The technique we outline in this paper seeks to
answer that question. Moreover, the method is quite general and may be
useful in other applications both inside and outside of neuroscience.

\subsection{Outline}
\Secref{background} reviews the necessary theory of
persistent homology at the level of elementary applied topology,
recapping also some of the major results that we use. We then briefly
describe the model we use for neuron activity.

\Secref{hidden} describes the process of inferring away the
contributions of covariates until the list of suspected such is empty,
whereupon information about the homology of any hidden covariates is
revealed. Demonstrations of the efficacy of the technique with
simulated neuronal data under various conditions follow in
\secref{results}.

Finally, \secref{generalized} sketches some possible
applications of our technique outside of neuroscience. We also attempt
to place it on a firmer and more general mathematical ground.

\subsection{Related work}
\citeauthor{curto2008cell}~\cite{curto2008cell} were the first to
employ tools from algebraic topology to study neural data in general,
and in particular to attempt reconstruction of the homology of an
animal's physical environment from such data.

\citeauthor{dabaghian2012}~\cite{dabaghian2012} form a model of neuron
activity, and use persistent homology very much along the lines
sketched above to study qualitative properties of this model. The
properties include the time taken to form a homologically correct
representation of the environment, and the robustness of this
representation with regard to model
parameters. \citeauthor{dabaghian2014}~\cite{dabaghian2014} improve on
this model, add theta wave phase-precession, and study its effect on
the persistent homology learning of the environment.

\citeauthor{giusti2015}~\cite{giusti2015} very recently showed how
persistent homology can help determine whether the information encoded
in the spike trains of neurons encodes for something \emph{geometric},
or is just random, thus answering our question~Q1.

Our work is in part based on the same approach as~\cite{giusti2015},
but seeks also to answer questions~Q0 and Q2 within one framework, and
should also be applicable in a very general setting not specific to
neuroscience.

\section{Background} \label{sec:background}

In this section we survey the necessary tools from persistent
homology in particular and topological data analysis (TDA) in general,
and fix basic notation. Thorough introductions can be found
in~\cite{ghrist2014, edelsbrunner2014, frogbook}. Familiarity with
elementary algebraic topology is assumed (see~\cite{hatcher} for an
introduction).

The $p$-simplices of a simplicial complex $K$ with vertex set $V$ will
be written like $[i_0, i_1, \dotsc i_p]\subseteq K$, where each
$i_0,\dotsc,i_p\in V$ are implicitly distinct. The geometric
realization of $K$ will be denoted $|K|$. Unless otherwise is noted,
$H_k(K)$ denotes the $k$'th simplicial homology of $K$, computed with
coefficients in $\mathbb{Z}/2\mathbb{Z}$. All simplicial complexes we
consider are assumed to be finite.

As hinted at in \secref{intro}, our basic data will be
\emph{intersection information}, or a proxy thereof. The following
basic construction that lies at the heart of TDA encodes such data as
a simplicial complex.
\begin{definition}
Let $X$ be a topological space. The \defword{nerve} of a collection of
sets $\mathcal{U} = \{U_i\subseteq X \suchthat i \in I\}$ is the
simplicial complex $N\mathcal U$ defined by
  \begin{equation*}
    [i_0, i_1, \dotsc, i_p] \subseteq N \mathcal U \iff
    U_{i_0}\cap U_{i_1} \cap \dotsm \cap U_{i_p} \neq \emptyset.
  \end{equation*}
\end{definition}
%\colorbox{red}{\textbf{FIXME}: legge til en figur her? }

The nerve construction's importance to TDA stems from its ability to
discretely encode the homotopy type of a topological space, as the
\emph{nerve theorem} shows. The theorem exists in a more general form,
but for the purpose of our paper it suffices to state it for metric
spaces.
\begin{theorem} \label{thm:nerve}
  Let $X$ be a metric space, and let $\mathcal U = \{U_i \suchthat i
  \in I\}$ be a finite closed cover of $X$ with the property that for
  all subsets $J\subseteq I$, the intersection $\bigcap_{j\in J}U_j$
  is either empty or contractible. Then $X$ and $|N\mathcal U|$ have
  the same homotopy type.
%  \begin{equation*}
%    X \hty |N\mathcal U|.
%  \end{equation*}
\end{theorem}
A cover satisfying the above condition is said to be \defword{good}.
Throughout, we let $B(x; r)$ denote the closed ball of radius $r$
centered at $x\in \RR^n$. One aspect of TDA is the study of point
clouds --- finite point sets in Euclidean space --- by applying
\thmref{nerve} to a nerve of closed balls.
\begin{definition}
  Let $P=\{p_1, p_2, \dotsc, p_N\}\subseteq\RR^n$ be a point
  cloud. The \defword{\Cech{} complex} of $P$ at \defword{scale}
  $\varepsilon$ is the nerve
  \begin{equation*}
    \Ccplx(P; \varepsilon) = N\{ B(p; \varepsilon/2) \suchthat
    p \in P \}.
  \end{equation*}
  We write $[i_0,\dotsc,i_k]\subseteq \Ccplx(P; \varepsilon)$ for the
  $k$-simplex corresponding to the intersection $B(p_{i_0};
  \varepsilon) \cap \dotsm \cap B(p_{i_k}; \varepsilon)$.
\end{definition}

Since a finite set of closed balls is a good cover of its union,
\thmref{nerve} ensures that the purely combinatorial nerve reflects
the homotopy type of that union. Thus, with $H^{\text{sing}}_\ast$
denoting singular homology with coefficients in $\ZZ/2\ZZ$,
\begin{equation}
  H^{\text{sing}}_\ast\left(\bigcup_{p\in P} B(p;
  \varepsilon/2)\right) \iso 
  H^{\text{sing}}_\ast\left(\lvert\Ccplx(P;
  \varepsilon)\rvert\right) \iso
  H_\ast\left(\Ccplx(P;
  \varepsilon)\right). \label{eq:fundamental}
\end{equation}
In \eqref{fundamental}, the first isomorphism is a consequence
\thmref{nerve}, while the second is a standard result in algebraic
topology. Computing the right hand side amounts to doing linear
algebra.

\subsection{Persistent homology} \label{sec:ph}
If a point cloud is assumed to have been sampled from some unknown
subspace $X$ of $\mathbb{R}^n$, then it is reasonable to apply
\eqref{fundamental} in order to reconstruct
$H^{\text{sing}}_{\ast}(X)$. This of course begs the question: what is
the ``correct'' scale at which to view the point cloud, or, explicitly
in the case of the \Cech{} complex, what (if any) is the ``right''
ball radius?  Persistent homology sidesteps the question by
ecompassing \emph{all} scales in one unifying construction.

We define the necessary constructions in a rather compact language
below. The unfamiliar reader may want to read the wordier
survey~\cite{ghrist2008barcodes}.

\begin{definition}
  A \defword{persistence module $\mathbf{V}$ (over $\mathbb{R}$)} is a
  collection of finite-dimensional $\mathbf{k}$-vector spaces
  $\{\mathbf{V}(t) \suchthat t\in \mathbb{R}\}$ and linear maps
  $\{\mathbf{V}(s\leq t): \mathbf{V}(s)\to \mathbf{V}(t) \suchthat
  s,t\in \mathbb{R}, s\leq t\}$ satisfying $\mathbf{V}(s\leq s) =
  {\id}_{\mathbf{V}(s)}$ and $\mathbf{V}(s'\leq t)\circ
  \mathbf{V}(s\leq s') = \mathbf{V}(s\leq t)$ for all $s\leq s'\leq
  t$.
\end{definition}
In other words, $\mathbf{V}$ is functor from $\mathbb{R}$ to finite
dimensional $\mathbf{k}$-vector spaces. We define an \emph{interval}
$I\neq \emptyset\subseteq \mathbb{R}$ to be a set such that if $s,
t\in I$, then $s'\in I$ for all $s'$ satisfying $s\leq s'\leq t$.

\begin{definition} 
  The \defword{interval persistence module} $\boldchi_I$ on an
  interval $I\subseteq \RR$ is given by
  \begin{align*}
    \boldchi_I(s) = \begin{cases}
      \mathbf{k} &\text{ if } s\in I \\
      0          &\text{ otherwise}    \end{cases}
  \end{align*}
  where any two nontrivial vector spaces are connected by the identity
  map.
\end{definition}
Direct sums of persistence modules are defined point-wise, \ie
$(\mathbf V \oplus \mathbf W)(s) = \mathbf V(s) \oplus \mathbf W (s)$
and similarly for the maps, and those that can only be written as
trivial direct sums are called \defword{indecomposable}. It is easy to
see that interval persistence modules are indecomposable. An existence
theorem~\cite{crawley2015decomposition} guarantees that there for
every $\mathbf{V}$ exists a multiset $B(\mathbf{V})$ of intervals in
$\mathbb{R}$ such that
\begin{equation*}
  \mathbf{V} \iso
  \bigoplus_{I\in B(\mathbf{V})}\boldchi_{I}.
\end{equation*}
That this decomposition is essentially unique is ensured by the
Azumaya--Krull--Remak--Schmidt theorem~\cite{azumaya1950}. In
particular, (the isomorphism class of) a persistence module
$\mathbf{V}$ is completely determined by the multiset
$B(\mathbf{V})$. This is the famous \defword{barcode} description of
$\mathbf{V}$. Moreover, an interval in $\mathbb{R}$ can be canonically
identified with a point in $(\RR\cup \{\pm \infty\})^2$ by means of
its endpoints. Hence, a persistence module is completely described by
a multiset of points in $(\RR\cup \{\pm \infty\})^2$; we shall refer
to this description as the \defword{persistence diagram} of
$\mathbf{V}$.

\begin{definition}
  A finite filtration $K$ of simplicial complexes
  \begin{equation*}
    K_{\varepsilon_0} \subseteq K_{\varepsilon_1} \subseteq \dotsm
    \subseteq K_{\varepsilon_N}
  \end{equation*}
  for $-\infty = \varepsilon_0<\varepsilon_1<\dotsm<\varepsilon_N$ gives rise to the $k$'th
  \defword{persistent homology module $PH_k(K)$ of $K$} defined by
  \begin{equation*}
    PH_k(K)(s) = H_k(K_{\max\{\varepsilon_i \suchthat \varepsilon_i \leq s\}})
  \end{equation*}
  with the linear maps induced by inclusions. This is commonly known
  as a \defword{sublevel} persistence module.
\end{definition}
We will often refer to $PH_\ast(K)$ as a persistent homology module
without specifying the homology dimension. Observe that as $K$ is a
finite filtration of simplicial complexes, there will only be a
finite number of intervals in $B(PH_*(K))$.

It is often useful to reduce the information content of persistent
homology to an even simpler, if incomplete, descriptor. In line with
the $k$'th Betti number being defined as the rank of the $k$'th
homology group, we define the following.
\begin{definition}
  Let $\mathbf{V}$ be a persistence module. Its \defword{Betti curve}
  is the function $\beta:\RR\to\NN$ defined by $\beta(s) = \dim
  \left(\mathbf{V}(s)\right)$. When referring to persistent homology
  modules $PH_\ast(K)$, we write $\beta_k$ for the Betti curve of
  $PH_k(K)$.
\end{definition}
There are obviously non-isomorphic persistence modules that give rise
to the same Betti curve, so this descriptor discards information. Its
advantage is that, as functions, Betti curves are easy to compare,
especially if considered as bounded functions $\RR\to\RR$. This
becomes particularly useful if the persistence diagrams contain a
great number of points. See also Sections~\ref{sec:comparepers} and
\ref{sec:random}.

\subsection{Some filtered simplicial complexes} \label{sec:somecomplexes}
Finite filtrations of simplicial complexes abound in TDA. One natural
example is the \Cech{} filtration
\begin{equation*}
  \Ccplx(P; \varepsilon_1) \subseteq \Ccplx(P; \varepsilon_2)
  \subseteq \dotsm \subseteq \Ccplx(P; \varepsilon_N)
\end{equation*}
of a finite point cloud $P$ given a sequence of radii
$\varepsilon_1<\varepsilon_2<\dotsm<\varepsilon_N$. We write
$\Ccplx(P)$ for the filtration.

The \Cech{} filtration of a point cloud is of theoretical importance
since \thmref{nerve} applies not only to each of its individual
complexes (as in \eqref{fundamental}), but also functorially
so~\cite{chazal2008towards} (respecting maps). In many applications,
including the one in this paper, the data we are given do not consist
of points in Euclidean space, but rather of relationships (such as
distances) between points in an unknown metric space. Many simplicial
complexes can be built from such data, for example on
graphs~\cite{jonsson2008}.
\begin{definition}
  Let $G=(V,E)$ be a graph. The \defword{flag complex} or
  \defword{clique complex} of $G$ is the largest simplicial complex
  $\flag(G)$ whose $1$-skeleton is $E$.
\end{definition}

If a graph $G=(V,E)$ has edge weights $w:E\to\RR$, defining
\begin{equation*}
  G_\varepsilon = (V,w\inv((-\infty, \varepsilon]))
\end{equation*}
gives a filtration of flag complexes
\begin{equation*}
  V \subseteq \flag(G_{\varepsilon_1}) \subseteq \flag(G_{\varepsilon_2})
  \subseteq \dotsm \subseteq \flag(G_{\varepsilon_N})
\end{equation*}
for a sequence $\varepsilon_1<\varepsilon_2<\dotsm<\varepsilon_N$.

The terms ``flag complex'' and ``\Cech{} complex'', and their
corresponding notation, will also interchangeably refer to the
filtered versions.

Flag complexes can be constructed, and be useful, for any underlying
graph. However, they play a special role when that graph is a metric
space. Indeed, if $P=\{p_1,\dotsc, p_N\}\subseteq\RR^n$ is a point
cloud, then the complete graph $G$ on the vertices $\{1,\dotsc,N\}$,
with edge weights
\begin{equation*}
  w(i,j) = \lVert p_i-p_j\rVert_2
\end{equation*}
gives a (filtration of) flag complex(es) called the
\defword{Vietoris--Rips complex} of $P$. We write $\VR(P)$ for the
entire filtration and $\VR(P;\varepsilon)$ for the complex at scale
$\varepsilon$ in the filtration.

We obviously have the inclusion
$\Ccplx(P;\varepsilon)\subseteq\VR(P;\varepsilon)$ at any scale
$\varepsilon$. Conversely, if the points $p_0,\dotsc, p_k\in P$ are
arranged at the vertices of the standard $k$-simplex in Euclidean
space and scaled so that they are $\varepsilon$ apart pairwise, then
$[0,1,\dotsc, k] \subseteq \Ccplx(P; \sqrt{2}\varepsilon)$. The
standard simplices are in fact the worst cases~\cite{frogbook}, and so
\begin{equation*}
  \Ccplx(P;\varepsilon) \subseteq \VR(P;\varepsilon) \subseteq \Ccplx(P;\sqrt{2}\varepsilon)
\end{equation*}
for any scale $\varepsilon$. This sandwiching of $\Ccplx(P)$ and
$\VR(P)$ implies that homological features that persist under the
morphism $H_k(\VR(P;\varepsilon)) \to H_k(\VR(P;
\sqrt{2}\varepsilon))$ correspond to features of $H_k(\Ccplx(P);
\varepsilon))$, and vice versa. The persistent homology of the
Vietoris--Rips complex thus contains important information from that
of the \Cech{} complex, while using only distance data.

\subsection{Comparing persistence modules} \label{sec:comparepers}
We now define a family of metrics on multisets in
$(\RR\cup\{\pm\infty\})^2$, which in turn pull back to (extended)
(pseudo)metrics on persistence modules. Such metrics allow us to
compare persistence modules with one another in a quantitative way.

\begin{definition}
  Let $A$ and $A'$ be finite multisets in
  $(\RR\cup\{\pm\infty\})^2$, \ie persistence diagrams, and let $\Delta$
  denote the diagonal with countably infinite multiplicity at every
  point. Write $M$ for the set of all multiset bijections
  $A\cup\Delta\to A'\cup\Delta$. For $q=1,2,\dotsc,\infty$, let
  $d^{\mathrm{E}}_q$ denote the extended Euclidean $q$-metric on
  $(\RR\cup\{\infty\})^2$. Then for $p=1,2,\dotsc$ define the
  \defword{Vaser\v{s}te\u{\i}n distance}
  \begin{equation*}
    d_{p,q}(A,A') = \inf_{f\in M}\left(\sum_{a\in
      A}\left(d^{\mathrm{E}}_q(a, f(a)\right)^p\right)^{1/p}
  \end{equation*}
  and for $p=\infty$ the \defword{bottleneck distance}
  \begin{equation*}
    d_{\infty,q}(A,A') = \inf_{f\in M}\sup_{a\in A} d^{\mathrm{E}}_q(a, f(a)).
  \end{equation*}
\end{definition}

The computation of these distances is equivalent to a maximum bipartite matching problem, and so can be performed in $\mathcal{O}\left(\max(|A|, |A'|)^3\right)$ time using the
(improved) Kuhn--Munkres algorithm~\cite{kuhn1955,munkres1957}. For the purpose
of our paper we will only consider $p=1$ and $q=2$. 

We remark that the bottleneck distance $d_{\infty,q}$ is of great
theoretical importance because of its role in \emph{stability
  theorems}~\cite{cohensteiner2007stability}.  For example, if $P'$ is
an $\varepsilon$-perturbation of a point cloud $P$ in Euclidean space,
then $d_{\infty, \infty}(PH_\ast(\Ccplx(P)),
PH_\ast(\Ccplx(P')))\leq\varepsilon$ and $d_{\infty,
  \infty}(PH_\ast(\VR(P)), PH_\ast(\VR(P')))\leq \varepsilon$. In
particular, persistent homology is \emph{stable} with respect to
perturbation of the input data.

\subsection{Statistics of persistence modules and Betti curves} \label{sec:random}
Later in the paper, we will build filtered simplicial complexes from
measurements, and we may wonder how persistent homology depends on the
statistical properties of the measurements.

The survey of
\citeauthor{kahle2014randomsurvey}~\cite{kahle2014randomsurvey} covers
much of what is known about some special cases of random simplicial
complexes, namely those that are flag complexes of Erdős--Rényi random
graphs\footnote{An $(n,p)$-\defword{Erdős--Rényi} random graph has $n$
  nodes, and each possible edge appears independently with probability
  $p$. We will refer to flag complexes of such graphs as \defword{ER
    complexes} with parameters $n$ and $p$.} and those that are
Vietoris--Rips complexes of random points from Euclidean space (the
latter will hereafter be referred to as \defword{random geometric
  complexes}). While most results known are asymptotic in the number
of vertices in the complex, and thus of little direct relevance to our
setting, some qualitative conclusions can be drawn also about the
finite case assuming the the size of the vertex set is not too small.

\begin{figure}[htbp]
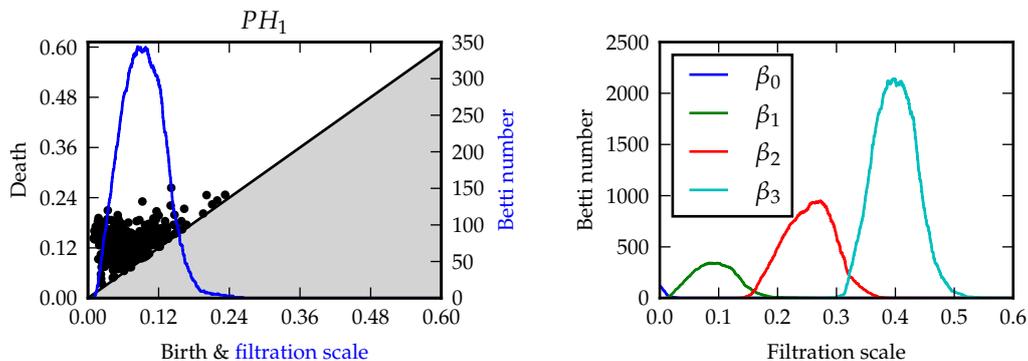

  \centering
  \fastfig{random}{er-120_pd_1_60}\fastfig{random}{er-120_betti_60}
  \caption{Persistent homology of a realization of an Erdős--Rényi
    complex with $120$ vertices. Increasing homology dimension results
    in an increase in peak Betti number and a narrowing of the edge
    probability range where non-vanishing Betti numbers are
    likely. This behavior is a signature of ER
    complexes.} \label{fig:random_er}
\end{figure}

\begin{figure}[htbp]
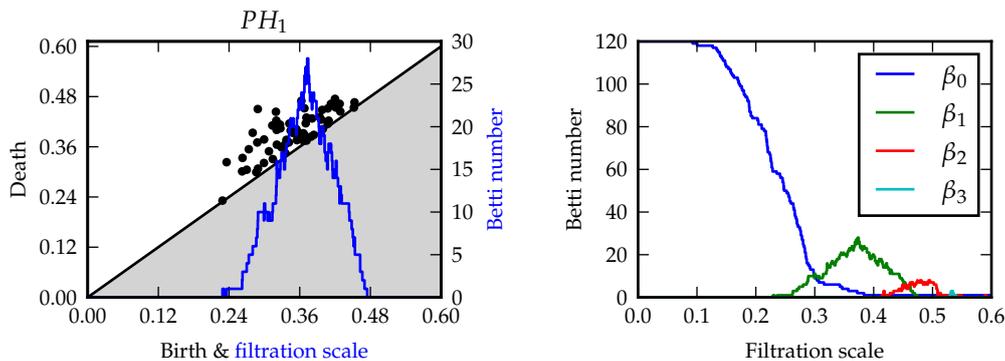

  \centering
  \fastfig{random}{geometric-120_pd_1_60}\fastfig{random}{geometric-120_betti_60}
  \caption{Persistent homology of a random geometric complex, the
    Vietoris--Rips complex of $120$ points sampled uniformly at random
    from the unit cube in $\RR^4$.} \label{fig:random_geometric}
\end{figure}

The first qualitative observation is that non-bounding cycles, \ie
homology generators, are likely to occur only in ER complexes when the
edge probability parameter is in a certain range, and this range
becomes narrower as the homology dimension grows. Within the ``allowed
range'', however, a large Betti number may occur. As the homology
dimension increases, the allowed range shrinks while the peak Betti
number within the range grows. \Figref{random_er} illustrates
the behavior.
The second observation is that in the random \emph{geometric} setting,
large Betti numbers are much less likely. Intuitively, the triangle
inequality puts heavy constraints on where points may be placed in
space for the cycles they are a part of not to become bounding at
comparatively small filtration scales. Highly persistent non-bounding
cycles are thus geometrically fragile and very sensitive to point
coordinates, and thus unlikely to occur by chance. As homology
dimension increases, the peak Betti number goes down, as one would
expect from intuition, since high-dimensional cycles are even more
susceptible to being filled in by the triangle
inequality. \Figref{random_geometric} illustrates the
behavior.

The takeaway for us is that the Betti curves of ER complexes behave
substantially different from geometric complexes and, of course, from
those of complexes arising from structured data. Betti curves can
therefore act as an indicator of the absence of meaningful
information; contrast \figref{random_er} and
\figref{random_geometric}. Experiments show that around $100$ vertices
seems to be more than sufficient for this indicator to be robust.

\subsubsection{Quantitatively testing for randomness} \label{sec:test_for_random}
If $G$ is a complete weighted graph on $V$, we write $\shuf(G)$ for (a
realization of) the complete graph on $V$ having the edge weights of
$G$ randomly shuffled. Except in degenerate cases $\flag(\shuf(G))$
should be a good realization of an ER random complex. The discussion
above then suggests two ad hoc measures for how consistent a flag
complex is with an ER random complex by comparing $PH_\ast(\flag(G))$
and $PH_\ast(\flag(\shuf(G)))$.

For various $p$ and $q$ we define
\begin{equation*}
  \delta_{k}(G) = d_{p,q}\left(PH_k(\flag(G)),
  PH_k(\flag(\shuf(G)))\right),
\end{equation*}
\ie the persistence module metric between $PH(\flag(G))$ and its
shuffled version. We further let $\beta_k$ and $\beta'_k$ denote the
Betti curve of $PH_k(\flag(G))$ and of $PH_k(\flag(\shuf(G)))$,
respectively, and define the ratio
\begin{equation*}
  \Delta_k(G) = \frac{\max_s \beta_k(s)}{\max_s \beta'_k(s)}
\end{equation*}
whenever it exists. $\Delta_k(G)$ thus compares the peak Betti number
of $\flag(G)$ to that of its shuffled version. If, for as large a $k$
as is computationally feasible, $\delta_k(G)$ is close to zero and
$\Delta_k(G)$ is close to one, we have an indication that $\flag(G)$
resembles an ER random complex.

\subsection{Model for neuron activity}
We need a model for neuron activity for two reasons. Most importantly,
the process of inferring away the contribution that specific
covariates have on the neuron activity, as sketched in
\secref{intro} and detailed in \secref{hidden},
requires such a model. In addition, we prefer to work with synthetic
data so that we are in complete control of all ``experimental''
parameters when testing the feasibility of our technique.

It has recently been shown~\cite{roudi2011mean, roudi2015multi,
  dunn2015correlations} that statistical physics' \emph{kinetic Ising
  model}, a simple generalized linear model (GLM), does a good job
modelling networks of neurons.
%for non-equilibrium ferromagnetism also does a good job modelling
%networks of neurons.
In its original language, the model governs the discrete time
evolution of a set of spins under the influence of both an external
driving field and the neighboring spins (as defined either by a
discretization of Euclidean space in the original applications, or in
general by a weighted graph). In general, the model may be defined as
follows.
\begin{definition} \label{def:ising}
  A set of $\pm 1$-valued random variables $S_i(k)$, with $1\leq i
  \leq N$ and $k=1,2,\dotsc$, are said to obey the (discrete time)
  \emph{kinetic Ising model} with \emph{couplings} $J\in\RR^{N\times
    N}$ and \emph{external fields} $E_1,\dotsc,E_n:\NN\to\RR$ if their
  conditional probabilities are
  \begin{align*}
    P\left(S_i(k+1)=s_i(k+1) \given S_1(k) = s_1(k), \dotsc, S_N(k) = s_N(k)\right) \\
    = \frac{\exp\left(s_i(k+1)\left(E_i(k) + \sum_{j=1}^N
      J_{i,j}s_j(k)\right)\right)}{2\cosh\left(E_i(k) + \sum_{j=1}^N
      J_{i,j}s_j(k)\right)}.
  \end{align*}
  For ease of notation, we set $s_1(0)=\dotsm=s_N(0)=-1$. We refer to
  the expression $F_i(k)=E_i(k) + \sum_{j=1}^N J_{i,j}s_j(k)$ as the
  system's \emph{Hamiltonian} (at time step $k$).
\end{definition}

In the neuroscience interpretation of the model, the probability of
neuron $i$ firing or not at time step $k$, as signified by $S_i(k)$
taking the value $+1$ or $-1$ respectively, is governed by the
external field $E_i(k)$ and by whether or not the neighboring neurons
$\{j \suchthat J_{i,j}\neq 0\}$ fired during time step $k-1$. We will
in \secref{hidden} make vital assumptions about the external
fields $E_1,\dotsc,E_N$.

\begin{figure}[htbp]
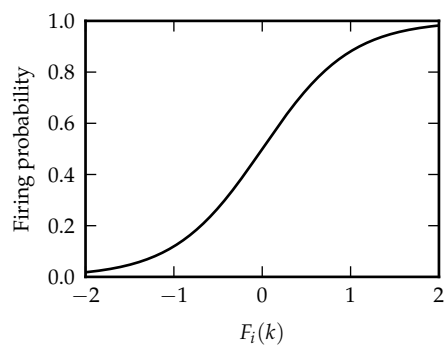

  \centering
  \fastfig{sigmoidal}{sigmoidal_60}
  \caption{Firing probability as a function of the Hamiltonian in the
    GLM.} \label{fig:sigmoidal}
\end{figure}

The function $x\mapsto \exp(x)/\cosh(x)$ is a sigmoidal, and
\figref{sigmoidal} shows the firing probability of a cell (\ie the
probability of $s=1$ in \defref{ising}) as a function of the
Hamiltonian.

\begin{figure}[htbp]
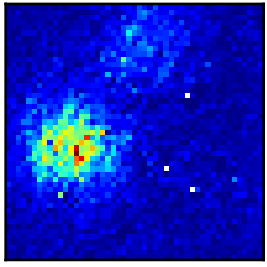

  \centering
  \fastfig{ising-place-fields}{cell-24_60}
  \caption{Example of the place field of one cell in a population of
    $100$ whose activities are governed by the GLM. Colors have
    qualitatively the same meaning as in
    \figref{placefield}.} \label{fig:ising-placefield}
\end{figure}

\Figref{ising-placefield} shows the place field of a cell whose firing
is generated by the GLM. In the terminology of \defref{ising}, the
external field is in this case predominantly a spatial Gaussian,
together with a smaller head direction tuning Gaussian. The cells are
randomly, but sparsely, coupled.

\section{Uncovering hidden information} \label{sec:hidden}
We suppose that a neuroscientist provides a list of \emph{candidate
  stimuli}, such as for example spatial preference, head direction
preference, theta phase preference, and so forth. We further suppose
that the influence of such stimuli can be described by certain
functions of distances on some simple manifolds. We will
specificically consider
\begin{itemize}
\item boxes $(0,1)^d$ of any dimension $d$, with the Euclidean metric,
\item boxes with any number number of $d$-disks removed so long as
  these do not disconnect the manifold,
\item circles, with their usual Riemannian metric,
\item spheres, with their usual Riemannian metric,
\item products of the above, with the product metric.
\end{itemize}
If the candidate list contains $L$ stimuli with corresponding
manifolds $M_1,\dotsc, M_L$ from the list above, then a complete
description of the relevant state of the animal at any time is assumed
to be a point in the product $M = M_1 \times \dotsm \times M_L$. An
experiment is then performed wherein the neurons are recorded together
with samples of the trajectory $\alpha:\RR\to M$ of the animal through
its state space.

The neuronal activity recorded is temporally binned with width $\delta
t$ (in this paper $\delta t\approx 10$ ms). Thus the primary input
data to our method consist of a spike train for each neuron, \ie $N$
binary vectors of the form
\begin{equation*}
  s_i = (s_i(1), s_i(2), \dotsc, s_i(T)) \in \{-1,1\}^T,
\end{equation*}
together with samples of the state space trajectory
\begin{equation*}
  \alpha(1), \alpha(2), \dotsc, \alpha(T) \in M.
\end{equation*}

As a concrete example, we might imagine a hypothetical situation where
the researcher believes that only spatial position and head direction
govern neuron activity. In this case, an experiment might be performed
wherein a rat explores a box, while its position and the direction of
its head are recorded at the same regular time instances as neuron
activity. Supposing that $100$ neurons are recorded over the course of
$10$ minutes\footnote{This is a realistically sized data set according
  to computational neuroscience data sharing website
  \url{http://crcns.org}.}, the data we are given then consist of
$100$ binary spike train vectors of length $60000$ together with
$60000$ samples of the rat state as points in the state space
$(0,1)^2\times \mathbb{S}^1$. Our goal, as outlined in
\secref{contrib}, is then to determine whether the two suspected
stimuli --- spatial position and head direction --- describe the
observed activity, and, most importantly, if not, what the homological
properties of any remaining unknown stimuli are.

\subsection{Spike train order complex}
Having obtained the necessary spike trains, we compute a measure of
the degree of cofiring between all pairs of neurons. The Pearson
correlation of the corresponding spike trains is a sensible choice,
and for $s\in\RR^{T}$ and $s'\in\RR^{T'}$ we thus define
\begin{equation}
  \corr(s, s') = \frac{\sum_{k=1}^{\min(T,T')}
    (s(k)-\overbar{s})(s'(k)-\overbar{s}')}
       {\sqrt{\sum_{k=1}^{\min(T,T')}
           (s(k)-\overbar{s})^2}\sqrt{\sum_{k=1}^{\min(T,T')}
           (s'(k)-\overbar{s}')^2}}, \label{eq:pearson}
\end{equation}
where
\begin{equation*}
  \overbar{s} = \frac{1}{\min(T,T')}\sum_{k=1}^{\min(T,T')}s(k).
\end{equation*}

To reduce the Pearson correlation's sensitivity to errors in the
binning process and slight systematic timing errors in measurements,
we will average over a small number of time bins. For a vector
$s=\left(s(1), \dotsc, s(T)\right)\in\RR^T$, we will write its $i$'th
left shift as
\begin{equation*}
  s[i] = \left(s(1+i), \dotsc, s(T)\right) \in \RR^{T-i}.
\end{equation*}
Then for $\tau \geq 0$, define
\begin{equation*}
  \corravg_\tau(s, s') = \frac{1}{\tau+1}\max\left(\sum_{i=0}^\tau
  \corr(s[i], s'), \sum_{i=0}^\tau \corr(s, s'[i])\right).
\end{equation*}

Finally, to follow the convention that simplicial complexes are
filtered in order of \emph{increasing} distance, we define for a set
of spike trains $s_1,\dotsc, s_N$ the \defword{spike train distances}
\begin{equation*}
%  D(s_i, s_j) = 1 - \corr(s_i, s_j). % Remember that we must reintroduce the max if we take back delays.
  D_\tau(s_i, s_j) = \max_{k,l\in\{1,\dotsc, N\}}\corravg_\tau(s_k, s_l) - \corravg_\tau(s_i, s_j).
\end{equation*}
We will throughout this paper write $D=D_0$ and only briefly discuss
specific choices of the averaging time scale $\tau$ in
\secref{results}.
%% With the above data in hand, we need to compute a measure of the
%% degree of cofiring between two spike trains. We shall let $\langle s,
%% s\prime \rangle$ denote the inner product of vectors $s$ and $s\prime$, and
%% have $s[n]$ denote the left-shifted sequence
%% \begin{equation*}
%%   s[n] = (s(n), s(n+1), \dotsc, s(T-1), 0, \dotsc, 0) \in \{0,1\}^{T}.
%% \end{equation*}
%% For a fixed \emph{correlation range} $\tau\geq 0$, we define the
%% \emph{cross-correlation} of two spike trains $s$ and $s\prime$ by
%% \begin{equation*}
%%   \xcorr(s,s\prime) = \max\left( \sum_{n=0}^\tau \ip{s}{s\prime[n]},
%%   \sum_{n=0}^\tau \ip{s[n]}{s\prime} \right).
%% \end{equation*}
%% To follow the convention that simplicial complexes are filtered in
%% order of increasing distance, we may for a set of spike trains
%% $s_1,\dotsc, s_N$ define the spike train distances

While $D_\tau$ is not a metric when $\tau\neq 0$ (clearly the triangle
inequality need not hold), it does provide some semblance of a
closeness notion for spike trains, and should be a useful indirect
measure of how much place fields intersect. An obvious next step is
therefore to take the flag complex of the graph with vertices $1,
\dotsc, N$ corresponding to the neurons, and whose edge weights are
$w(i,j) = D(s_i, s_j)$. However, as pointed out in~\cite{giusti2015},
the chemical and biological processes that go into a neuron building
its action potential and firing, and the physical processes that
constitute the measurement of that event, are of a highly non-linear
nature. What we observe, then, is merely a highly transformed version
of the real cofiring relationship of neurons. In other words, if we
think of $C(i,j)$ as a true low-level measure of the cofiring
relationship of neurons $i$ and $j$, then $D(s_i,s_j) = f(C(i,j))$ for
some unknown function $f:\RR\to\RR$. On biological grounds it is
reasonable to assume that the only thing that is known, and can ever
be known, about $f$ is that it is monotonic.

Following~\cite{giusti2015}, we rescale our data so as not to
prescribe meaning to the actual values of the distance $D$, since it
may well be the case that no such meaning exists.
\begin{definition}
  Let $G=(E,V)$ be the complete graph on the vertices $V=\{1,\dotsc,
  n\}$ with edge weights $w$. Let $\varphi:\{1,\dotsc,\binom{n}{2}\}\to
  E$ be a bijection that sorts the weights
    \begin{equation*}
      w(\varphi(1)) < w(\varphi(2)) < \dotsm
    \end{equation*}
    (breaking ties arbitrarily), and let $\widetilde{G}$ be the
    complete graph on $V$ with the edge weights
    \begin{equation*}
      \widetilde{w}(i,j) = \frac{\varphi^{-1}(i,j)}{\binom{n}{2}}.
    \end{equation*}
    The \defword{order complex} of $G$ is the flag complex $\OC(G) =
    \flag(\widetilde{G})$.
\end{definition}
If we take the order complex of the spike trains and their distances,
we discard all information except precisely that which survives the
unknown monotone transformation $f$, namely the ordering of the edge
weights.

\subsection{Removing the contribution of stimuli to reveal hidden information} \label{sec:remove}
The likelihood of data
\begin{equation*}
  \mathbf{s} = \{s_i(k) \suchthat 1\leq i \leq N, 1\leq k\leq T\}
\end{equation*}
observed under the GLM is
\begin{equation*}
  L(\mathbf{s}) =
  \prod_{i=1}^N\prod_{k=1}^{T-1}\frac{\exp\left(s_i(k+1)\left(E_i(k) +
    \sum_{j=1}^N J_{i,j}s_j(k)\right)\right)}{2\cosh\left(E_i(k) +
    \sum_{j=1}^N J_{i,j}s_j(k)\right)}.
\end{equation*}
We now assume that the external field part of the Hamiltonian
decomposes into a sum of Gaussians on the various factors of the state
space $M$ corresponding to candidate covariates.

As before, $M = M_1 \times \dotsm \times M_L$, where each $M_i$ is
assumed to be one of the manifolds listed earlier
(\secref{generalized} discusses a
generalization). Projections onto the factors are written $\pr_i:M\to
M_i$, and we denote by $d_i$ the metric on $M_i$. We define the
Gaussians $V_{l,q}:M_l \to \RR$ by
\begin{equation*}
  V_{l,q}(x) = \exp\left(-\frac{\left(d_l(x, c_{l,q})\right)^2}{2\sigma_{l,q}^2}\right)
\end{equation*}
for $1\leq l \leq L$ and $1\leq q \leq Q_l$, and assume that the
external part of the Hamiltonian for neuron $i$ at time step $k$ can
be written (recall that $\alpha$ is the animal's path through its
state space)
\begin{equation*}
  E_i(k) = \sum_{l=1}^L\sum_{q=1}^{Q_l} A_{i,l,q}\left(V_{l,q}\circ\pr_l\circ\alpha\right)(k)
\end{equation*}
for $c_{l,q}\in M_l$, $A_{i,l,q}\in\RR$ and
$\sigma_{l,q}\in\RR$.

We may use the language of \secref{intro} if for each $i$ and $l$
there is only one $q$ for which $A_{i,l,q}\neq 0$. Fix an $i$ and an
$l$ and let $q$ be the only index for which $A_{i,l,q}\neq 0$. Then
$c_{l,q}$ is the center of a place field corresponding to covariate
number $l$, $\sigma_{l,q}$ is a measure of its width, while
$A_{i,l,q}$ specifies the peak strength with which it influences the
firing of neuron $i$. In other words, neuron $i$ has a place field
specified by $c_{l,q}$ and $\sigma_{l,q}$. When the above relationship
between the indices of $A_{\bullet,l,\bullet}$ does not hold, we allow
each covariate to govern place cell activity through a linear
combination of Gaussian fields, which will be necessary for the
inference process described next.

The $\log$-likelihood of the observed data is
\begin{align}
  \log(L(\mathbf{s})) = \sum_{i=1}^N\sum_{k=1}^{T-1}&\Bigg[
  s_i(k+1)\left(\sum_{l=1}^L\sum_{q=1}^{Q_l} A_{i,l,q}\left(V_{l,q} \circ \pr_l \circ \alpha\right)(k) + \sum_{j=1}^N J_{i,j}s_j(k)\right) \nonumber\\
  &-\log\left(2\cosh\left( \sum_{l=1}^L\sum_{q=1}^{Q_l} A_{i,l,q}\left(V_{l,q} \circ \pr_l \circ \alpha\right)(k) + \sum_{j=1}^N J_{i,j}s_j(k)  \right)\right)
  \Bigg], \label{eq:loglikelihood}
\end{align}
and it is an easy calculus exercise to show that
\begin{align*}
  \begin{cases}
    \RR^{N\sum_{l=1}^L Q_l} \to \RR \\
  (A_{\bullet,\bullet,\bullet}, J_{\bullet,\bullet}) \mapsto \log(L(\mathbf{s}))
  \end{cases}
\end{align*}
is a convex function. We can therefore perform likelihood maximization
by means of convex optimization to infer the $A_{i,l,q}$'s and
$J_{i,j}$'s that best fit the observed data. With these coefficients
in hand, we can selectively remove the (expected) contribution of each
stimulus on the candidate list. The residual data after removal no
longer consists of binary spike trains, but instead real-valued time
series. Low and high values correspond to improbable and probable
firing events, respectively, while values near zero reflect lack of
knowledge of the probability of either outcome.

Throughout this paper all the inference will be performed with
$25^2=625$ spatial basis functions, and $25$ circular basis functions,
both with means uniformly distributed. For example, if the state space
is $M=M_1\times M_2$ with $M_1=(0,1)^2$ and $M_2 = \mathbb{S}^1$, then
$Q_1=25^2$ and $Q_2=25$, and with the $c_{1,q}$'s forming a regular
grid on $(0,1)^2$ and the $c_{2,q}$'s uniformly distributed on
$\mathbb{S}^1$.

\begin{figure}[htbp]
  \centering
      \begin{tikzpicture}[font=\small]
      %\node () at (0,7) {a};
      %\node () at (10,0) {b};

      \node[draw,align=center,rounded corners,anchor=south east] (source) at (0,0) {Experiment\\ or\\ simulation};
      \node[draw,align=center,parallelogram,anchor=east] (spiketrains) at ($ (source.north west) - (1,0) $) {Spike\\ trains};
      \draw[-{>[sep=1pt]}] (source.west) to [bend left=10] (spiketrains.bottom right corner);

      \node[draw,align=center, rectangle, anchor=east] (ph) at ($ (spiketrains.west) - (1,0) $) {Persistent\\ homology};
      \draw[-{>[sep=1pt]}] (spiketrains.west) to (ph.east);

      \node[draw,align=center, parallelogram, anchor=south] (barcodes) at ($ (ph.north) + (0,0.5) $) {Persistence diagrams\\ and Betti curves};
      \draw[-{>[sep=1pt]}] (ph.north) to (barcodes.south);

      \node[draw,align=center, diamond, anchor=south, shape aspect=3] (israndom) at ($ (barcodes.north) + (-1,0.5) $) {Random?};
      \draw[-{>[sep=1pt]}] (barcodes.north) to (israndom.south);

      \node[draw,align=center, rounded corners, anchor=east] (random) at ($ (israndom.east) + (1,1) $) {No more\\ to learn};
      \draw[-{>[sep=1pt]}] (israndom.east) to [bend right=30] node[below,sloped,midway]{Yes} (random.south);

      \node[draw,align=center, diamond, shape aspect=3, anchor=south, inner sep=1pt] (isexhausted) at ($ (israndom.north) + (0,1) $) {Stimuli list exhausted?};
      \draw[-{>[sep=1pt]}] (israndom.north) to node[above,sloped,midway]{No} (isexhausted.south);

      \node[draw,align=center, rounded corners, anchor=east] (info) at ($ (israndom.west) - (0.2,1) $) {Information\\ about\\ topology of\\ hidden stimuli};
      \draw[-{>[sep=1pt]}] (isexhausted.west) to [bend right=30] node[above,sloped,midway]{Yes} (info.north);
      \draw[-{>[sep=1pt]}, dashed] (barcodes.west) to [bend left=30] (info.east);

      \node[draw,rectangle, anchor=north west,text width=5cm] (inference) at ($ (isexhausted.east) + (1,0) $) {
        \begin{enumerate}[label=\roman*)]
          \item Pick stimulus; \Eg spatial position, head direction, theta phase preference, $\dotsc$
          \item Infer its contribution to spike trains
          \item Remove contribution
        \end{enumerate}
      };
      \draw[-{>[sep=1pt]}] (isexhausted.east) to node[above,midway,sloped]{No} (inference.north west);
      \draw[-{>[sep=1pt]}] (inference.south) to [bend left=25] (spiketrains.top right corner);

%      \visible <2-> {
%        \node[anchor=south west] () at (inference.south west) {\includegraphics[width=4cm]{figures/params_number0732_fakedata_exp0000_R02620_002000_maps_0005.pdf}};
%        }
    \end{tikzpicture}
  \caption{Our main contribution summarized.} \label{fig:flowchart}
\end{figure}
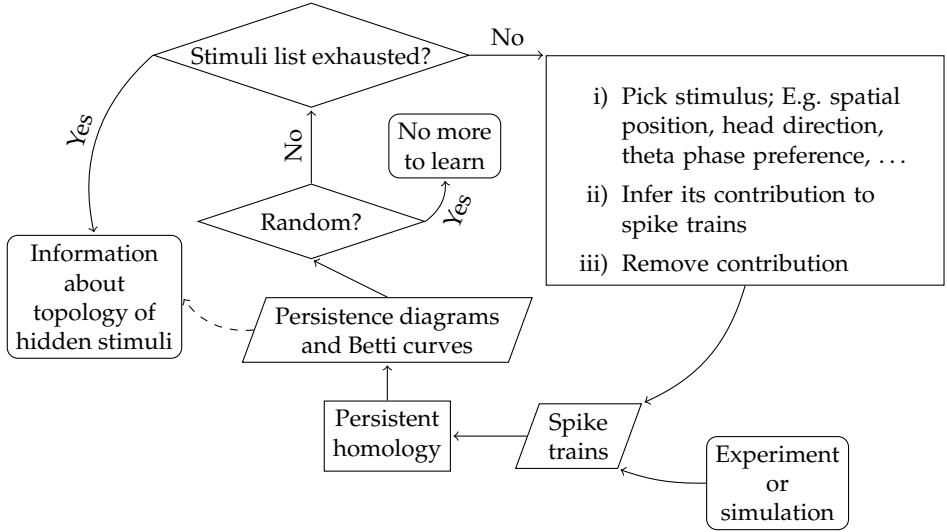

The assumptions on the external fields above thus allow us to remove
the contributions of specific stimuli towards the firing activity,
suggesting the following algorithm:
\begin{enumerate}
\item Perform an experiment as described above, yielding spike trains
  $s_1,\dotsc,s_N$ and samples of the state space path $\alpha$. \label{step:exp}
\item Compute the spike train distances $D(s_i, s_j)$ for $1\leq
  i,j\leq N$, and let them be the edge weights on a complete graph $G$
  on $N$ vertices. \label{step:corr}
\item Compute $PH_\ast(\OC(G))$. \label{step:ph}
\item Is $PH_\ast(\OC(G))$ consistent with the persistent homology of
  an ER random complex per \secref{test_for_random}?  If so,
  go to step~\ref{step:random_yes}. If not, go to
  step~\ref{step:exhausted}.
  \begin{enumerate}
    \item There is nothing more to learn from our method. We are done. \label{step:random_yes}
  \end{enumerate}
\item Is the list of candidate stimuli exhausted? If so, go to
  step~\ref{step:exhausted_yes}. If not, go to
  step~\ref{step:remove}. \label{step:exhausted}
  \begin{enumerate}
    \item $PH_\ast(\OC(G))$ may reveal information about the homology
      of the state space corresponding to unknown covariate(s). We are
      done. \label{step:exhausted_yes}
  \end{enumerate}
  \item Using the inference process described above, remove the
    contributions of a selected stimulus. This yields real-valued
    time-series that from now on replace the spike trains $s_1,\dotsc,
    s_N$. Go to step~\ref{step:corr}. \label{step:remove}
\end{enumerate}
\Figref{flowchart} summarizes the above procedure.

As a concrete illustration, we return to the example from earlier:
assume that place cell activity is in fact governed only by spatial
position and head direction, but suppose that head direction tuning is
unknown to the researchers. The candidate covariate list therefore
consists only of spatial position. For an experiment conducted with
the animal exploring a box, the state space is $(0,1)^2$. After having
performed steps~\ref{step:corr} through~\ref{step:remove} once, the
second iteration leads us to step~\ref{step:exhausted_yes}; there are
no more stimuli to account for, yet we see persistent homology
inconsistent with random data. Examination of the persistence diagrams
of $PH_\ast(\OC(G))$ reveals that the unknown covariates correspond to a
state space with the same homology as a circle. Together with other
evidence, this may lead the researchers to suspect head direction
tuning as a hidden covariate. New experiments may then be performed to
investigate this, and our method may be applied also to the new data.

We finally point out that it is \emph{not} essential to our method
that we necessarily capture the correct homology of the full state
space. It is the state space after the removal of known stimuli that
matters (and even then, the weaker information of whether the
remaining state space is homologically trivial or not may be useful).

\section{Computational results} \label{sec:results}
We test the efficacy of our method using synthetically generated data
in order to be in complete control of all ``experimental parameters'',
and because publicly available real data often come from experiments
without a topological focus (for example the spatial environment tends
to be homologically trivial). For the results presented here, neuron
activity was simulated from the same GLM as used for the inference
process by appropriate selection of peak field strength coefficients
$A_{\bullet,\bullet,\bullet}$, centers $c_{\bullet,\bullet}$ and
widths $\sigma_{\bullet,\bullet}$. In addition, a constant negative
term (typically $-1$) was added to the external fields to make the
overall firing rate low outside of place fields, as is the case for
many real cells. This term essentially just lowers the noise floor of
our data, and should be of no deeper significance to us.

Even if one of the experiments below does not call for spatial or head
direction tuning, the state space will always consist of at least a
factor for the physical environment and a factor $\mathbb{S}^1$ for
the head direction. If the physical environment is denoted $B$, then
the $B\times\mathbb{S}^1$ factor of the state space is explored as
follows: If at some time step the animal's state is $(x,y,\theta)\in
B\times\mathbb{S}^1$, then the next state is found by choosing a
$\theta'$ randomly and uniformly within $0.02$ of $\theta$ in
$\mathbb{S}^1$. If $(x + 5\cdot 10^{-4}\cos\theta', y+5\cdot
10^{-4}\sin\theta', \theta')$ is within $B\times\mathbb{S}^1$,
this point is the next state. If not, new angles are drawn until the
new state is valid. We note that qualitatively similar results are
obtained if this slightly realistic random walk is replaced by
ordinary Brownian motion or uniform random sampling.

Except in \secref{exp_int}, we do not average the correlation
measure, \ie we work with the spike train distance $D_0$.

Some additional computational experiments have been relegated to
Appendix~\ref{sec:extra_experiments}.

\subsection{Recovering spatial homology} \label{sec:exp_simple_recover}
While the central point of this paper is the uncovering of the
homological properties of unknown stimuli underlying neural activity,
we begin by providing an example of how persistent homology of the
order complex of spike train distances can recover the homology of a
spatial environment. This is analogous to some of the results
presented in~\cite{dabaghian2012}.

State space is now the unit square punctured by four disks of radius
$0.15$. The disks are centered at $(0.27, 0.27)$, $(0.27, 0.72)$,
$(0.72, 0.27)$ and $(0.72, 0.72)$. We refer to the punctured box as
$B$ below. For technical reasons, data were generated with the state
space having an extra circle factor, but every cell had its activity
coefficients corresponding to this factor set to zero. Thus \emph{only
  spatial tuning} is present in this example.

In the notation of \secref{remove}, $N=Q_1=Q_2=100$, $M_1 =
B$, $M_2=\mathbb{S}^1$, $L=2$, and $J_{i,j}=0$ for all $1\leq i, j\leq
N$. The $c_{1,q}$'s form a regular  grid on $B$, and
\begin{align*}
  A_{i,1,q} = \begin{cases}
    2 &\text{ if } i=q \\
    0 &\text{ otherwise},
  \end{cases}
\end{align*}
while $A_{i, 2, q} = 0$ for all
$i,q$. Figure~\ref{subfig:simple-recover/45_space} shows the
spatial tuning of a single neuron.

\begin{figure}[htbp]
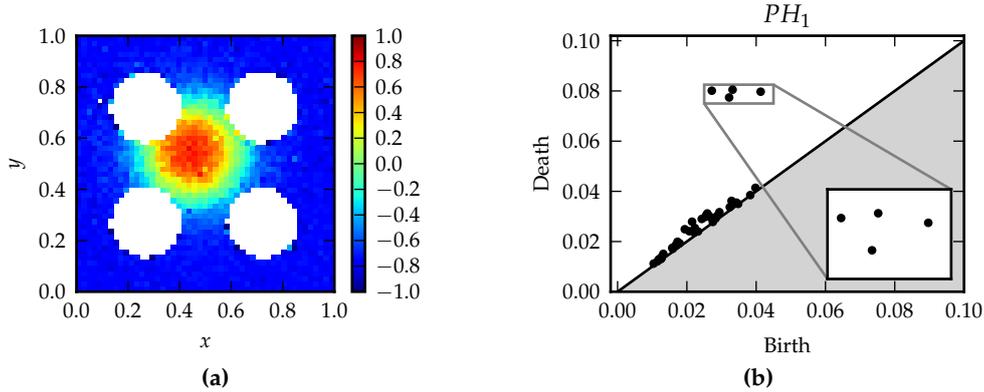

  \centering
  \begin{subfigure}[b]{0.48\textwidth}
    \centering
    \renewcommand{\localfigname}{45_space}
    \renewcommand{\localfigprefix}{simple-recover}
    \fastfig{\localfigprefix}{\localfigname_60}
    \caption{} \label{subfig:\localfigprefix/\localfigname}
  \end{subfigure}
  \begin{subfigure}[b]{0.48\textwidth}
    \centering
    \renewcommand{\localfigname}{cPsa_ordered_pd_1}
    \renewcommand{\localfigprefix}{simple-recover}
    \fastfig{\localfigprefix}{\localfigname_60}
    \caption{} \label{subfig:\localfigprefix/\localfigname}
  \end{subfigure}
  \caption{Results of the experiment in \secref{exp_simple_recover}.
    \mysubref{subfig:simple-recover/45_space} The activity of a
    single cell as a function of spatial
    position. \mysubref{subfig:simple-recover/cPsa_ordered_pd_1}
    Persistent homology of the order complex. All higher dimensions of
    homology are trivial or close to
    trivial.} \label{fig:exp_simple_recover}
\end{figure}

Figure~\ref{subfig:simple-recover/cPsa_ordered_pd_1} shows
that we correctly recover the homology of $B$, the only part of $M$
detected by the neural activity. In other words, this example
illustrates how we can detect and even count the obstructions in a
space from neuron activity alone.

\subsection{Proof of concept} \label{sec:exp_poc}
The simplest possible setting wherein our inference scheme, laid out
in \secref{remove}, is useful, is perhaps one where both
spatial and head direction tuning govern neuron activity, but where
the researcher believes only one of those to be real.

For this computation, the spatial component of the state space is a
unit square punctured in its center by a single disk of radius $0.2$,
denoted by $B$. The head direction component is $\mathbb{S}^1$.

In the notation of \secref{remove}, $N=Q_1=Q_2=100$, $M_1=B$,
$M_2=\mathbb{S}^1$, $L=2$, and $J_{i,j}=0$ for all $1\leq i,j\leq
N$. The $c_{1,q}$'s form a regular grid on $B$, and
\begin{align*}
  A_{i,1,q} = \begin{cases}
    2 &\text{ if } i=q \\
    0 &\text{ otherwise},
  \end{cases}
\end{align*}
Similarly, the $c_{2,q}$'s are uniformly spread out over
$\mathbb{S}^1$. To avoid artificially coupling head direction and
spatial tuning through the ordering of their place field centers, we
let $\sigma$ be a random permutation of $\{1,\dotsc, N\}$ and then let
\begin{align*}
  A_{i,2,q} = \begin{cases}
    2 &\text{ if } i=\sigma(q) \\
    0 &\text{ otherwise},
  \end{cases}
\end{align*}
for all $i,q$.

We reiterate that we assume that the researcher is \emph{unaware of
  head direction tuning} as a real influence on place cell activity in
this example; he believes spatial position is the only relevant
stimulus. After conducting an experiment, the researcher sees the
neurons' spatial dependence examplified in
Figure~\ref{subfig:exp_poc_pre_act_a}. The head direction dependence
in Figure~\ref{subfig:exp_poc_pre_act_b} is \emph{not} known to the
researcher.
\begin{figure}[htbp]
  \centering
  \begin{subfigure}[b]{0.45\textwidth}
    \fastfig{poc}{27_space_60}
    \caption{} \label{subfig:exp_poc_pre_act_a}
  \end{subfigure}
    \begin{subfigure}[b]{0.45\textwidth}
      \fastfig{poc}{27_head_60}
    \caption{} \label{subfig:exp_poc_pre_act_b}
  \end{subfigure}
  \caption{Activity of a single neuron in the experiment from
    \secref{exp_poc}. \mysubref{subfig:exp_poc_pre_act_a}
    Spatial dependence. This is the only view of activity visible to
    the researcher. \mysubref{subfig:exp_poc_pre_act_b} Head
    direction dependence. The researcher is \emph{not} privy to this
    information, as he is unaware that head direction preference is
    real.} \label{fig:exp_poc_pre_act}
\end{figure}

\begin{figure}[htbp]
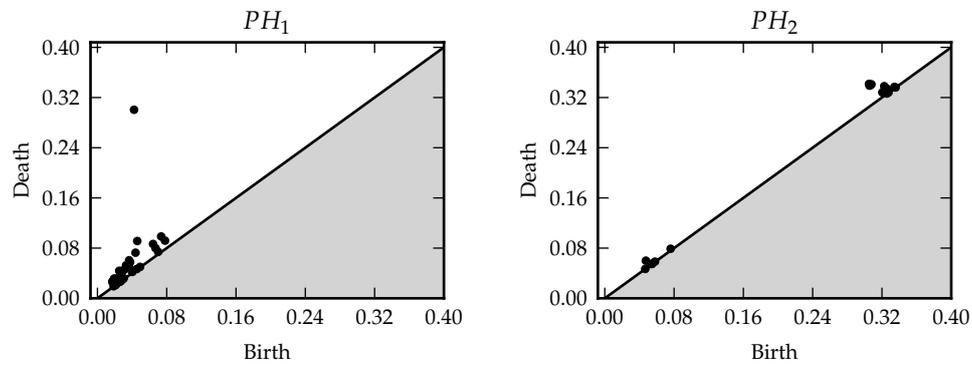

  \centering
  \fastfig{poc}{cPsa_ordered_pd_1_60}\fastfig{poc}{cPsa_ordered_pd_2_60}
  \caption{Observed persistent homology of the order complex of the
    correlations for the experiment from \secref{exp_poc}.} \label{fig:exp_poc_pre_pd}
\end{figure}

Persistent homology of the order complex of the correlations observed
can be seen in \figref{exp_poc_pre_pd}. Note that we do \emph{not}
observe the actual homology of the state space $M$, which is a
thickened torus. This is not entirely satisfactory, but at least the
observed persistence diagrams indicate there is homologically
nontrivial information present in the neuron activity.

Satisfied that the persistence diagrams are consistent with his
hypothesis about the relevant covariates, the researcher proceeds to
the next step, namely removing the effect of the spatial covariate
(the only one he is aware of). We maximize the expression in
\eqref{loglikelihood} in terms of the $A_{i,l,1}$'s using
L-BFGS-B\footnote{As provided by SciPy.}, and subtract from the
observed spike trains the expected activities predicted from only
spatial influence.

\begin{figure}[htbp]
  \centering
  \begin{subfigure}[b]{0.45\textwidth}
    \fastfig{poc}{27-c,sp_space_60}
    \caption{} \label{subfig:exp_poc_c,sp_act_a}
  \end{subfigure}
    \begin{subfigure}[b]{0.45\textwidth}
      \fastfig{poc}{27-c,sp_head_60}
    \caption{} \label{subfig:exp_poc_c,sp_act_b}
  \end{subfigure}
  \caption{Average activity of a single neuron in the experiment from
    \secref{exp_poc} \emph{after removal of spatial
      tuning}. \mysubref{subfig:exp_poc_c,sp_act_a} Spatial dependence. This is the only
    view of activity visible to the researcher. \mysubref{subfig:exp_poc_c,sp_act_b} Head
    direction dependence. The researcher is \emph{not} privy to this
    information, as he is unaware that head direction preference is
    real.} \label{fig:exp_poc_c,sp_act}
\end{figure}
\begin{figure}[htbp]
  \centering
  \fastfig{poc}{cPpa-c,sp_ordered_pd_1_60}\fastfig{poc}{cPpa-c,sp_ordered_pd_2_60}
  \caption{Observed persistent homology of the order complex of the
    correlations for the experiment from \secref{exp_poc}
    \emph{after removal of spatial tuning}.} \label{fig:exp_poc_c,sp_pd}
\end{figure}

The result of the removal on spatial activity tuning can be seen in
\figref{exp_poc_c,sp_act}. It is as expected, and obviously
does not provide new information to the researcher. The persistence
diagram in \figref{exp_poc_c,sp_pd}, however, shows that there
is still homologically non-trivial information contained in the
observed data. This should hopefully lead the researcher to suspect
that there are further, hidden, influences on neuron activity, and,
most importantly, that this/these \emph{are of a circular nature}.

Guided by this, the researcher might consider head direction
tuning. He therefore sets up a new experiment where also this is
tracked, so that also its influence may be removed from the
data. Doing so results in the activity plot in
\figref{exp_poc_c,sp,h_act}.
\begin{figure}[htbp]
  \centering
  \begin{subfigure}[b]{0.48\textwidth}
    \centering
    \renewcommand{\localfigname}{27-c,sp,h_space}
    \renewcommand{\localfigprefix}{poc}
    \fastfig{\localfigprefix}{\localfigname_60}
    \caption{} \label{subfig:\localfigprefix/\localfigname}
  \end{subfigure}
  \begin{subfigure}[b]{0.48\textwidth}
    \centering
    \renewcommand{\localfigname}{27-c,sp,h_head}
    \renewcommand{\localfigprefix}{poc}
    \fastfig{\localfigprefix}{\localfigname_60}
    \caption{} \label{subfig:\localfigprefix/\localfigname}
  \end{subfigure}
  \caption{Average activity of a single neuron in the experiment from
    \secref{exp_poc} \emph{after removal of both spatial and
      head direction tuning}. \mysubref{subfig:poc/27-c,sp,h_space} Spatial
    dependence. \mysubref{subfig:poc/27-c,sp,h_head} Head direction
    dependence.} \label{fig:exp_poc_c,sp,h_act}
\end{figure}
\begin{figure}[htbp]
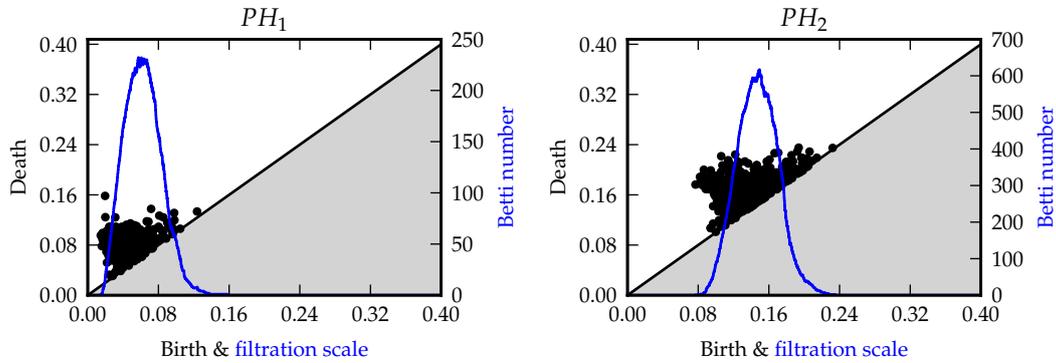

  \centering
  \fastfig{poc}{cPpa-c,sp,h_ordered_pd_1_60}\hspace{-1em}\fastfig{poc}{cPpa-c,sp,h_ordered_pd_2_60}
  \caption{Observed persistent homology of the order complex of the
    correlations for the experiment from \secref{exp_poc}
    \emph{after removal of both spatial and head direction
      tuning}.} \label{fig:exp_poc_c,sp,h_pd}
\end{figure}

The persistence diagrams in \figref{exp_poc_c,sp,h_pd} now
lack clear suggestions of non-trivial homology. Moreover, the Betti
curves are highly indicative of an ER random complex (compare
\figref{random_er}). To verify this last claim, we also
permuted the neuron correlations randomly;
\figref{exp_poc_c,sp,h_betti} shows the result. This should
serve as a firm indication that all stimuli underlying the neuron
activity in the data have been accounted for.
\begin{figure}[htbp]
  \centering
  \fastfig{poc}{cPpa-c,sp,h_betti_80}
  \caption{After all relevant covariates have been removed from the
    data in the experiment from \secref{exp_poc}, the Betti
    curves of the order complex of the correlations are consistent
    with those of an ER random complex (here formed by randomly
    shuffling the correlations). \Cf
    \figref{random_er}.} \label{fig:exp_poc_c,sp,h_betti}
\end{figure}

\subsubsection{Alternative scenario} \label{sec:exp_poc_alt}
One may also want to consider an alternative hypothetical scenario
wherein head direction tuning is the only suspected stimulus. For
readability reasons we do not include that scenario in full here. The
interesting part is the persistent homology after the removal of head
direction tuning. \Figref{exp_poc_c,h_pd} shows that we recover the
correct $PH_1$ also in this case.
\begin{figure}[htbp]
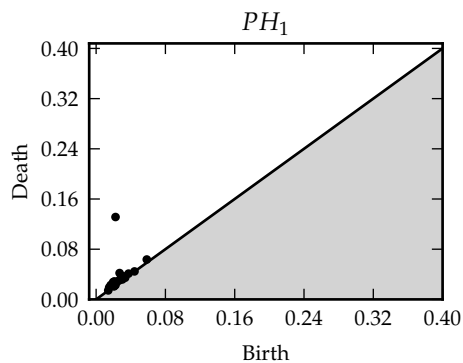

  \centering
  \fastfig{poc}{cPpa-c,h_ordered_pd_1_60}
  \caption{Persistent homology of the order complex after removing the
    effect if head direction tuning in an alternative version of the
    experiment in \secref{exp_poc} (see
    \secref{exp_poc_alt}).} \label{fig:exp_poc_c,h_pd}
\end{figure}

\subsection{Effect of couplings} \label{sec:exp_with-couplings}
In the preceding experiments, cells were never coupled. To illustrate
that our method also copes with such ``internal stimuli'', we repeated
the experiment from \secref{exp_poc} with the change that every cell
is given a weak but random coupling to a every other
cell. Specifically, we kept all simulation parameters as before, but
let every $J_{i,j}$ be drawn independently and uniformly from $[-0.1,
  0.1]$. The couplings are thus weak compared to the external stimuli
(which peak at $2$ in the centers of fields), but should nevertheless
introduce noise to the data.

\begin{figure}[htbp]
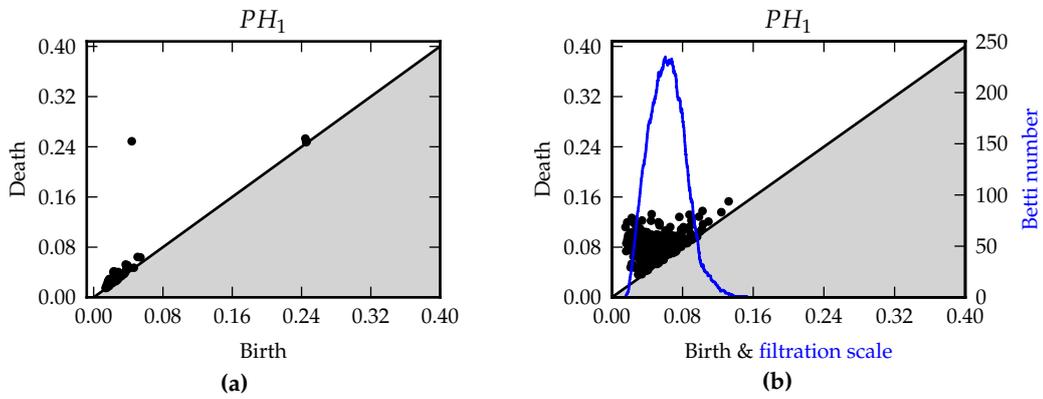

  \centering
  \begin{subfigure}[t]{0.48\textwidth}
    \centering
    \renewcommand{\localfigname}{cPpa-J,c,sp_ordered_pd_1}
    \renewcommand{\localfigprefix}{with-couplings}
    \fastfig{\localfigprefix}{\localfigname_60}
    \caption{} \label{subfig:\localfigprefix/\localfigname}
  \end{subfigure}
  \begin{subfigure}[t]{0.48\textwidth} % FIXME: The t is a dirty hack.
    \captionsetup{aboveskip=-1.75em} % FIXME: dirty hack.
    \centering
    \renewcommand{\localfigname}{cPpa-J,c,sp,h_ordered_pd_1}
    \renewcommand{\localfigprefix}{with-couplings}
    \fastfig{\localfigprefix}{\localfigname_60}
    \caption{} \label{subfig:\localfigprefix/\localfigname}
  \end{subfigure}
  \caption{Removing external and internal covariates in the experiment
    from \secref{exp_with-couplings}.
    \mysubref{subfig:with-couplings/cPpa-J,c,sp_ordered_pd_1}
    Persistence diagram after spatial dependence and internal
    couplings are removed.
    \mysubref{subfig:with-couplings/cPpa-J,c,sp,h_ordered_pd_1}
    Persistence diagram and Betti curve after all external (space and
    head) and internal (couplings) influences are
    removed.} \label{fig:exp_with-couplings_post}
\end{figure}

\Figref{exp_with-couplings_pre_act} shows the results before
any covariate removal. Observe that the random couplings introduce
significant noise in the spatial dependence of the activity compared
to that in
\figref{exp_poc_pre_act}. \Figref{exp_with-couplings_post}
shows that we are able to carry out the same procedure as in
\secref{exp_poc} also in the presence of couplings.
\begin{figure}[htbp]
  \centering
  \begin{subfigure}[b]{0.48\textwidth}
    \centering
    \renewcommand{\localfigname}{27_space}
    \renewcommand{\localfigprefix}{with-couplings}
    \fastfig{\localfigprefix}{\localfigname_60}
    \caption{} \label{subfig:\localfigprefix/\localfigname}
  \end{subfigure}
  \begin{subfigure}[b]{0.48\textwidth}
    \centering
    \renewcommand{\localfigname}{cPsa_ordered_pd_1}
    \renewcommand{\localfigprefix}{with-couplings}
    \fastfig{\localfigprefix}{\localfigname_60}
    \caption{} \label{subfig:\localfigprefix/\localfigname}
  \end{subfigure}
  \caption{Prior to removal of any covariates in the experiment from
    \secref{exp_with-couplings}. \mysubref{subfig:with-couplings/27_space}
    Spatial dependence for the activity of a single
    neuron. \mysubref{subfig:with-couplings/cPsa_ordered_pd_1}
    $PH_1$ persistence
    diagram.} \label{fig:exp_with-couplings_pre_act}
\end{figure}

\subsection{Effect of theta wave phase preference} \label{sec:exp_theta}
We simplistically model theta wave phase preference as each neuron
preferentially firing near a randomly chosen phase of a $7$ Hz
sinusoidal wave in time. The experimental parameters are the same as
in \secref{exp_poc}, except now $L=3$, and the state space
gains an extra factor $M_3=\mathbb{S}^1$. The $c_{3,q}$'s are
uniformly spread out over $\mathbb{S}^1$, and for a random
permutation\footnote{Present for the same reason as for the head
  direction tuning in \secref{exp_poc}.} $\sigma$ of
$\{1,\dotsc,N\}$ the field strength coefficients are
\begin{align*}
  A_{i,3,q} = \begin{cases}
    2 &\text{ if } i=\sigma(q) \\
    0 &\text{ otherwise},
  \end{cases}
\end{align*}
for all $i,q$.

\begin{figure}[htbp]
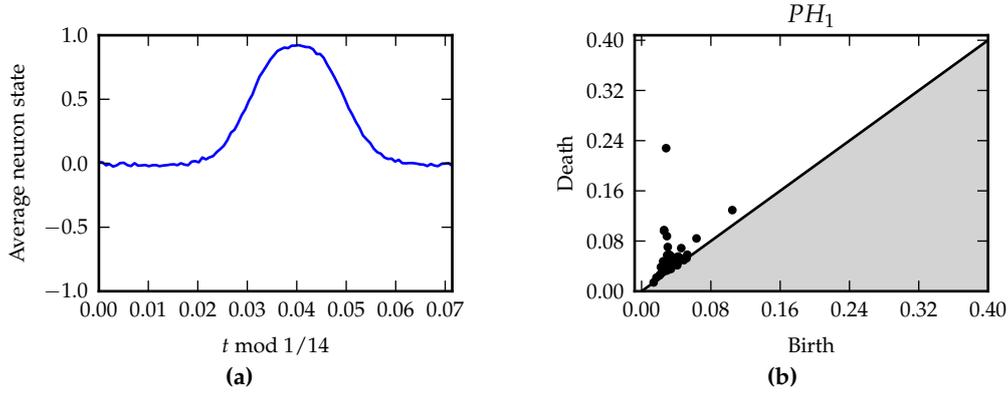

  \centering
  \begin{subfigure}[b]{0.48\textwidth}
    \centering
    \renewcommand{\localfigname}{27_theta}
    \renewcommand{\localfigprefix}{theta}
    \fastfig{\localfigprefix}{\localfigname_60}
    \caption{} \label{subfig:\localfigprefix/\localfigname}
  \end{subfigure}
    \begin{subfigure}[b]{0.48\textwidth}
    \centering
    \renewcommand{\localfigname}{cPsa_ordered_pd_1}
    \renewcommand{\localfigprefix}{theta}
    \fastfig{\localfigprefix}{\localfigname_60}
    \caption{} \label{subfig:\localfigprefix/\localfigname}
  \end{subfigure}

    \caption{Observations from the experiment in
      \secref{exp_theta} before removal of any covariates.
      \mysubref{subfig:theta/27_theta} Activity of a single
      neuron as a function of theta wave
      phase. \mysubref{subfig:theta/cPsa_ordered_pd_1}
      Persistent homology of the order complex of the spike train
      distances.} \label{fig:exp_theta_pre}
\end{figure}

Theta phase preference is thus, as far as topology is concerned,
precisely the same as head direction preference, except that the $M_3$
factor of state space is explored by always moving forward in time
(modulo $1/14$) instead of by a random walk. We therefore expect that
theta phase preference will contribute to homology in the same way as
head direction tuning. \Figref{exp_theta_pre} confirms this. Again it
should be pointed out that we are not observing homology consistent
with the three dimensional torus
$\mathbb{S}^1\times\mathbb{S}^1\times\mathbb{S}^1$. While this may
seem unsatisfactory, it is a quite natural effect of one covariate
suppressing the expression of the homology of the others, \ie one of
the radii of the torus dominating over the others. This illustrates
well why the inference process and sequential removing of covariates
really is necessary; the fact that there are \emph{three} circular
factors in the state space cannot be glanced directly from the
observed data.

\begin{figure}[htbp]
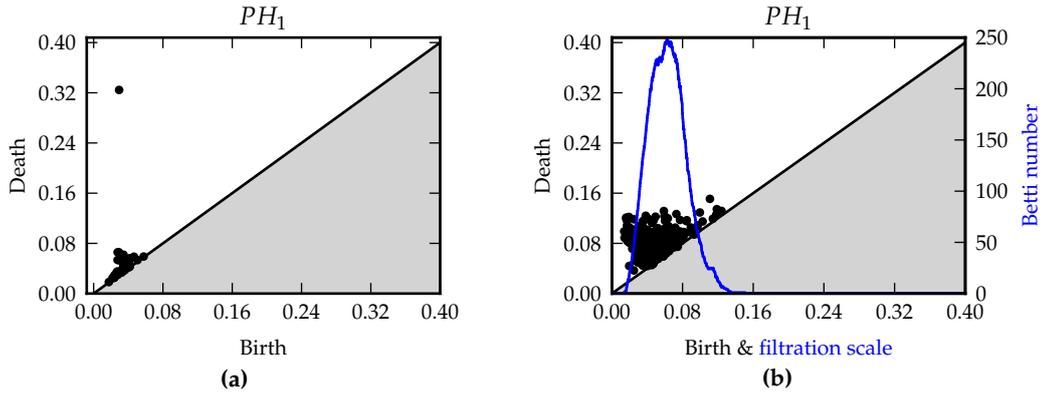

  \centering
  \begin{subfigure}[t]{0.48\textwidth}
    \centering
    \renewcommand{\localfigname}{cPpa-c,sp,h_ordered_pd_1}
    \renewcommand{\localfigprefix}{theta}
    \fastfig{\localfigprefix}{\localfigname_60}
    \caption{} \label{subfig:\localfigprefix/\localfigname}
  \end{subfigure}
  \begin{subfigure}[t]{0.48\textwidth} % FIXME: The t is a dirty hack.
    \captionsetup{aboveskip=-1.75em} % FIXME: dirty hack.
    \centering
    \renewcommand{\localfigname}{cPpa-c,sp,h,t_ordered_pd_1}
    \renewcommand{\localfigprefix}{theta}
    \fastfig{\localfigprefix}{\localfigname_60}
    \caption{} \label{subfig:\localfigprefix/\localfigname}
  \end{subfigure}
  
  \caption{Persistent homology of the order complex of the residual
    spike trains in the experiment in \secref{exp_theta}
    after removing some external
    covariates. \mysubref{subfig:theta/cPpa-c,sp,h_ordered_pd_1}
    After removing spatial and head direction
    tuning. \mysubref{subfig:theta/cPpa-c,sp,h,t_ordered_pd_1}
    After removing all covariates, including theta phase
    preference.} \label{fig:exp_theta_post}
\end{figure}

\Figref{exp_theta_post} shows that we obtain the expected results when
left with only theta phase preference and when all covariates are
removed.

\subsection{Sensitivity to field strength variations} \label{sec:exp_explore}
The preceding experiments illustrate the efficacy of our technique for
fixed choices of peak field strengths. We now wish to demonstrate that
our method is robust over a wide range of field strengths. To this
end, we repeat the experiment from \secref{exp_poc}, but now
for varying choices of peak strength for both the spatial and head
direction fields.

For conducting this experiment, we need a numerical measure of the
outcome that is more succinct than a persistence diagram or a Betti
cruve. Since we generate data with head direction and spatial tuning
enabled, we hope to see the same outcome as in the experiment from
\secref{exp_poc}:
\begin{itemize}
  \item Before removal of any covariates, we should see a prominent
    $PH_1$ generator, \ie one that stands out in the persistence
    diagram by having a longer lifetime than most. At the same time,
    the persistence diagrams and their associated Betti curves should
    be inconsistent with those of ER random complexes.
  \item This should remain the case when inferring away one of either
    head direction or spatial tuning.
  \item After inferring away both head direction and spatial tuning,
    the persistence diagrams and Betti curves should be highly
    consistent with those of ER random graphs.
\end{itemize}

As a very crude measure of the presence of ``highly persistent'' or
``prominent'' $PH_1$ generators, we simply consider the ratio of the
of the lifetime of the most persistent one to the lifetime of the
second-most persistent one. We refer to the ratio as $\rho_1$
throughout this section. Note that our aim is that $\rho_1$ measures
how easy a single prominent generator is to distinguish from noise. In
the case of multiple highly persistent $PH_1$ generators, such as for
the persistent homology of a torus, it reports a false
negative. However, since the experiment in \secref{exp_poc} failed to
capture this true homology of the state space, we feel confident this
measure is sufficient. We also verified this by manual inspection of
several of the persistence diagrams computed.

To compare with random ER complexes, we use both $\delta_1$ and
$\Delta_1$. Recall that the closer $\delta_1(G)$ is to zero and
$\Delta_1(G)$ is to one, the more consistent the flag complex built on
the data in $G$ is with an ER complex.

\begin{figure}[ptbh]
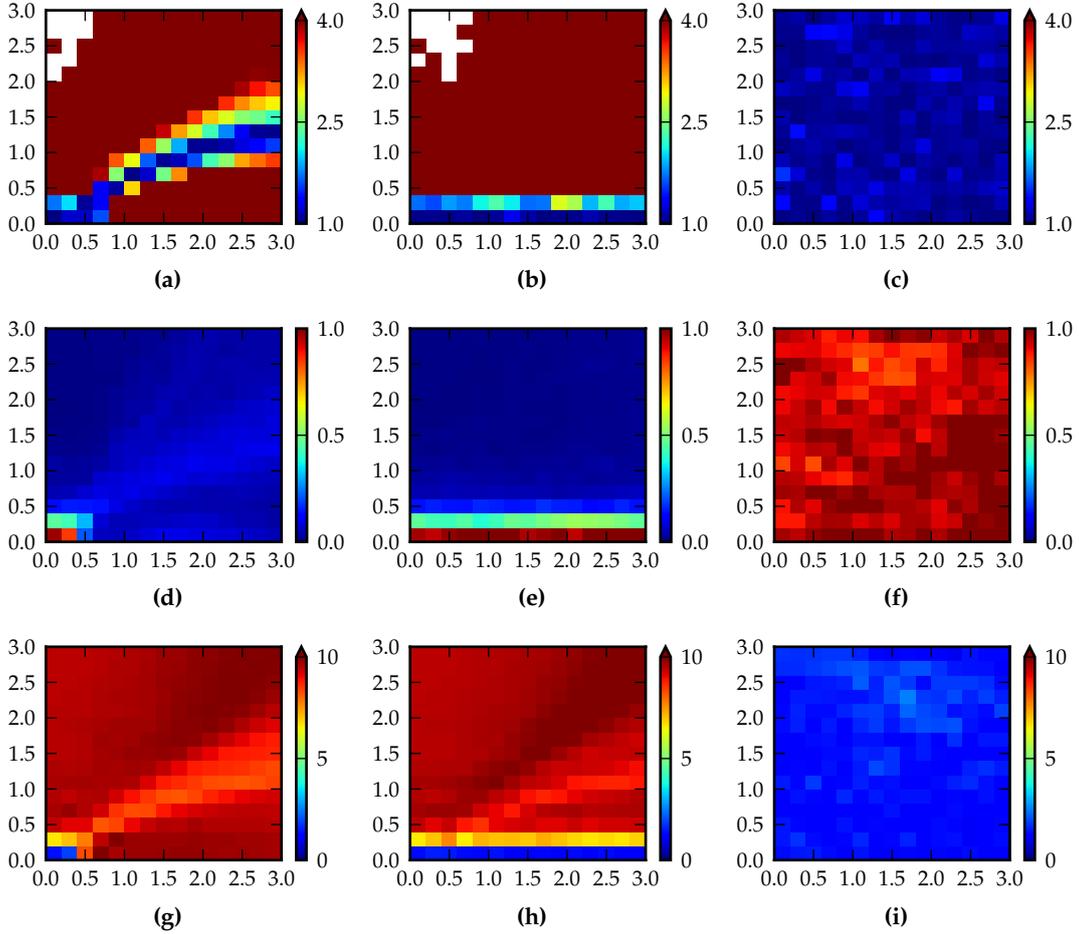

  \centering
  \begin{subfigure}[b]{0.32\textwidth}
    \captionsetup{aboveskip=-1.5em}
    \centering
    \renewcommand{\localfigname}{obs_longest_over_second}
    \renewcommand{\localfigprefix}{explore-success}
    \fastfig{\localfigprefix}{\localfigname_50}
    \caption{} \label{subfig:\localfigprefix/\localfigname}
  \end{subfigure}
  \begin{subfigure}[b]{0.32\textwidth}
    \captionsetup{aboveskip=-1.5em}
    \centering
    \renewcommand{\localfigname}{rem_c,sp_longest_over_second}
    \renewcommand{\localfigprefix}{explore-success}
    \fastfig{\localfigprefix}{\localfigname_50}
    \caption{} \label{subfig:\localfigprefix/\localfigname}
  \end{subfigure}
  \begin{subfigure}[b]{0.32\textwidth}
    \captionsetup{aboveskip=-1.5em}
    \centering
    \renewcommand{\localfigname}{rem_c,sp,h_longest_over_second}
    \renewcommand{\localfigprefix}{explore-success}
    \fastfig{\localfigprefix}{\localfigname_50}
    \caption{} \label{subfig:\localfigprefix/\localfigname}
  \end{subfigure}

  \begin{subfigure}[b]{0.32\textwidth}
    \captionsetup{aboveskip=-1.5em}
    \centering
    \renewcommand{\localfigname}{obs_peak_ratio}
    \renewcommand{\localfigprefix}{explore-success}
    \fastfig{\localfigprefix}{\localfigname_50}
    \caption{} \label{subfig:\localfigprefix/\localfigname}
  \end{subfigure}
  \begin{subfigure}[b]{0.32\textwidth}
    \captionsetup{aboveskip=-1.5em}
    \centering
    \renewcommand{\localfigname}{rem_c,sp_peak_ratio}
    \renewcommand{\localfigprefix}{explore-success}
    \fastfig{\localfigprefix}{\localfigname_50}
    \caption{} \label{subfig:\localfigprefix/\localfigname}
  \end{subfigure}
  \begin{subfigure}[b]{0.32\textwidth}
    \captionsetup{aboveskip=-1.5em}
    \centering
    \renewcommand{\localfigname}{rem_c,sp,h_peak_ratio}
    \renewcommand{\localfigprefix}{explore-success}
    \fastfig{\localfigprefix}{\localfigname_50}
    \caption{} \label{subfig:\localfigprefix/\localfigname}
  \end{subfigure}

  \begin{subfigure}[b]{0.32\textwidth}
    \captionsetup{aboveskip=-1.5em}
    \centering
    \renewcommand{\localfigname}{obs_pdm}
    \renewcommand{\localfigprefix}{explore-success}
    \fastfig{\localfigprefix}{\localfigname_50}
    \caption{} \label{subfig:\localfigprefix/\localfigname}
  \end{subfigure}
  \begin{subfigure}[b]{0.32\textwidth}
    \captionsetup{aboveskip=-1.5em}
    \centering
    \renewcommand{\localfigname}{rem_c,sp_pdm}
    \renewcommand{\localfigprefix}{explore-success}
    \fastfig{\localfigprefix}{\localfigname_50}
    \caption{} \label{subfig:\localfigprefix/\localfigname}
  \end{subfigure}
  \begin{subfigure}[b]{0.32\textwidth}
    \captionsetup{aboveskip=-1.5em}
    \centering
    \renewcommand{\localfigname}{rem_c,sp,h_pdm}
    \renewcommand{\localfigprefix}{explore-success}
    \fastfig{\localfigprefix}{\localfigname_50}
    \caption{} \label{subfig:\localfigprefix/\localfigname}
  \end{subfigure}

  \caption{Summary of the behavior of our method over varying field
    strengths, as detailed in \secref{exp_explore}. In all the plots,
    the horizontal axis denotes the peak spatial field strength and
    the vertical axis denotes the peak head direction field
    strength. Arrows in the color bar indicate that the colors are
    clipped above or below a value. For a spike train distance graph
    $G$ obtained using the given peak field strengths, the
    \textbf{first row} shows $\rho_1(G)$, the \textbf{second row}
    shows $\Delta_1(G)$, and the \textbf{third row} shows
    $\delta_1(G)$. In the \textbf{first column} $G$ comes from the
    original observations, in the \textbf{second column} the spatial
    influence is removed, and in the \textbf{third column} both
    spatial and head direction influence are
    removed.} \label{fig:exp_explore}
\end{figure}

\Figref{exp_explore} summarizes the relevant measures of
success. Observe that we meet our criteria for expected outcome for
parameters that lie outside the blue band in
\figref{exp_explore}\subref{subfig:explore-success/obs_longest_over_second}
and that at the same time are not too weak in either the spatial or
the head direction tuning strength (see \eg
\figref{exp_explore}\subref{subfig:explore-success/obs_pdm}). The
latter requirement is no surprise, as we with weak field strengths
observe neurons that barely respond to external stimuli. Further
investigation reveals that as the head direction field strength is
made weaker (\ie as we approach the band of failure from above in
\figref{exp_explore}\subref{subfig:explore-success/obs_longest_over_second}),
the most persistent $PH_1$ lives for a shorter and shorter time, until
it becomes indistinguishable from the noise near the diagonal. For
even weaker head direction field strengths, \ie below the band, we are
essentially in the domain of activity governed entirely by the spatial
fields.

Note also that when we in
Figure~\ref{subfig:explore-success/obs_longest_over_second}
are within the blue band,
Figures~\ref{subfig:explore-success/obs_peak_ratio} and
\ref{subfig:explore-success/obs_pdm} show that the observed
data are \emph{not} consistent with an ER complex. Inferring away the
spatial influence is therefore something one may still wish to do,
whereupon one uncovers homology strongly consistent with a circle
everywhere except for with weak head direction tuning, as seen in
Figure~\ref{subfig:explore-success/rem_c,sp_longest_over_second}.

\subsection{Homology from internal couplings} \label{sec:exp_int}
In the preceding experiments, internal couplings have been absent, or,
as in \secref{exp_with-couplings}, not themselves been the focus of
our attention. We now illustrate that our method is also capable of
detecting the homology of (the flag complex of) the graph defining the
neighbor relations of the neurons.

We generate data with only spatial fields and internal
couplings. Specifically, $N=Q_1=100$, $M=M_1=(0,1)^2$, the $c_{1,q}'s$ form
a regular grid on $(0,1)^2$, and the peak field strengths are
\begin{align*}
  A_{i,1,q} = \begin{cases}
    1 &\text{if } i=q \\
    0 &\text{othewise}
    \end{cases}
\end{align*}
(for technical reasons, just as in \secref{exp_poc}, the explored
state space still contains an extra $\mathbb{S}^1$ factor, but the
corresponding field strengths are set to zero). The symmetric matrix
$J$ describes a circle on all $N$ nodes with edges in both directions
with weights $2$. The indices defining the edges of the circle are
chosen randomly to avoid an unnatural coupling to the spatial fields
through the ordering.

\begin{figure}[htbp]
  \centering
  \begin{subfigure}[b]{0.48\textwidth}
    \centering
    \renewcommand{\localfigname}{cPsa1_ordered_pd_1}
    \renewcommand{\localfigprefix}{homology-from-couplings}
    \fastfig{\localfigprefix}{\localfigname_60}
    \caption{} \label{subfig:\localfigprefix/\localfigname}
  \end{subfigure}
  \begin{subfigure}[b]{0.48\textwidth}
    \centering
    \renewcommand{\localfigname}{cPsa10_ordered_pd_1}
    \renewcommand{\localfigprefix}{homology-from-couplings}
    \fastfig{\localfigprefix}{\localfigname_60}
    \caption{} \label{subfig:\localfigprefix/\localfigname}
  \end{subfigure}
  \caption{Persistent homology of the order complex of the observed
    spike train distances $D_\tau$ in the experiment from
    \secref{exp_int}. \mysubref{subfig:homology-from-couplings/cPsa1_ordered_pd_1}
    $\tau=1$. \mysubref{subfig:homology-from-couplings/cPsa10_ordered_pd_1}
    $\tau=10$.} \label{fig:exp_int}
\end{figure}

\Figref{exp_int}\subref{subfig:homology-from-couplings/cPsa1_ordered_pd_1}
shows an unexpected result consistent with (at least) \emph{two}
circles. Manual inspection of the spike train distances reveals that
this is an artifact of the GLM being coupled across only two
consecutive time steps. When we use a spike train distance without
temporal average (\ie $D_0$), we are unable to resolve any coupling
interactions across an even number of neurons. The $1$-skeleton of the
order complex of the spike train distances thus breaks into \emph{two}
circles, corresponding to the even and odd parity edges of the
neighborhood graph $(\{1,\dotsc, N\}, \{(i,j) \suchthat J_{i,j} \neq 0
\text{ or } J_{j,i} \neq 0\})$. This undesired behavior vanishes when
the correlation time average is greater, for example $\tau=10$, as is
shown in
Figure~\ref{subfig:homology-from-couplings/cPsa10_ordered_pd_1}.

\section{Discussion and sketches of a general framework} \label{sec:generalized}
We believe that the core aspects of the method presented in this paper
are applicable outside of neuroscience. In a general setting, we
imagine a point cloud $C=\{p_1, \dotsc, p_N\}$ of points on a manifold
$M$. The exact assumptions on $M$ have not been worked out, but we
believe that a compact, connected, homogenous Riemannian manifold
without boundary suffices.\footnote{Some of the state spaces
  considered in \secref{results} are of course not covered by these
  assumptions. Numerical evidence still strongly suggests that our
  method works well in practice also in the situations considered.}
Write the metric of $M$ as $d$. The points themselves, and their
pairwise distances, are a priori unknown. Estimates of the latter are
obtained by performing a random walk on $M$ while observing $N$
Bernoulli processes $S_1,\dotsc, S_N$ (producing time series
$s_1,\dotsc,s_N$). If we denote by $X(k)$ the random variable of the
random walk's position at time step $k$, the vital assumption is that
the parameter (the ``success'' probability) of Bernoulli process $S_i$
at time step $k$ is
\begin{equation}
  P\left(S_i(k) = 1 \given X(k)=x\right) = f\left(d(x,p_i)\right) \label{eq:gen_prob}
\end{equation}
for some monotonically decreasing sufficiently integrable (unknown)
function $f:\RR^{+}\to [0,1]$. If the random walk has progressed long
enough that the distribution of $X(k)$ is close to uniform, then we
are in a setting where we believe our methods are applicable.

The question to ask is whether the persistent homology of the flag
complex (or order complex) of the graph with edge weights $D_\tau(s_i,
s_j)$ for $1\leq i,j\leq N$ closely approximates the persistent
homology of $\VR(C)$.

In the above setup, define $g_k:M\times M\to\RR$ by
\begin{equation*}
  g_k(y,z) = \int_M f(d(y, x))f(d(z,x))\rmd \rho_k(x)
\end{equation*}
with $\rho_k$ the probability density function for the random walk at
time step $k$. The function $g_k$ arises naturally as the probability
\begin{equation*}
  P(S_i(k) = 1 \cap S_j(k) = 1) = g_k(p_i, p_j)
\end{equation*}
and is the essential part of the estimated Pearson correlation
$\corr(s_i, s_j)$ from \eqref{pearson}. In the case that $X(k)$ is
fully uniform, \ie $\mathrm{d}\rho_k=\mathrm{d}x$, proving correctness
of our recovered persistent homology amounts to showing the existence
of a monotonically decreasing $h:\RR^{+}\to\RR$ that makes
\begin{equation*}
  \begin{tikzpicture}
    \node (MxM) at (0,0) {$M\times M$};
    \node (R) at (2,0) {$\RR$}
     edge[<-] node[auto,swap] {$g$} (MxM);
    \node (Rp) at (0,-1) {$\RR^{+}$}
     edge[<-] node[auto] {$d$} (MxM)
     edge[->,dashed] node[auto,swap] {$h$} (R);
  \end{tikzpicture}
  % tikzcd does not play well with either TikZ externalize or the arXiv
  %% \begin{tikzcd}
  %%   M\times M   \arrow[d, "d"']  \arrow[r, "g"] & \RR \\
  %%   \RR^{+} \arrow[ur, dashed, "h"'] & \quad
  %% \end{tikzcd}
\end{equation*}
commute. In the Euclidean situation ($M=\RR^n$ and dropping the
assumptions on $M$) this is an easy calculus exercise, but the details
need to be worked out for the more general situations.

The actual persistent homology of (say, the Vietoris--Rips complex of)
$C$, \ie the level of question~Q0, may be of interest in applications
outside of neuroscience. While there are perhaps not many situations
in which observations of the Bernoulli processes are available, but
the values on the right hand side of \eqref{gen_prob} are \emph{not},
one example might be sensor network coverage
problems~\cite{ghrist2005coverage} with sensors of severely reduced
capabilities. Instead of being able to sense the strength of their
neighbors, we imagine that the sensors carry simple devices which
trigger with a frequency monotonically related to the distances to
other sensors.

Work should be done to give rigorous bounds on the persistence modules
based on the statistical properties of the Pearson correlations
$\corr(s_i, s_j)$, especially in the settings when the random walk
distribution is not yet truly uniform.

At the level of questions~Q1 and Q2 we encounter other systems
described by a GLM (such as in~\cite{borysov2015stock}) or, in
principle, by other models that accommodate an inference process like
that in \secref{remove}. Our method may be able to shed light on the
homological properties of parts of the external influences on such a
system, or, as in the experiment in \secref{exp_int}, unknown internal
couplings.

Future applications of our method on actual data should demonstrate
its usefulness in identifying hidden topological information in data,
both inside and outside of neuroscience.

%% We believe that our method should be applied to real data as soon as
%% these become available from experiments with non-trivial spatial
%% homology, or from experiments with other topologically interesting
%% covariates.

%\input{sketch.tex}
\section{Acknowledgments}
This paper is a product of a cooperative project where also Yasser
Roudi (Kavli Institute for Systems Neuroscience, Norwegian University
of Science and Technology) takes part.

GS would like to thank Geir-Arne Fuglstad for valuable discussions on
matters of probability.

All persistent homology computations were performed using Nanda's
\textit{Perseus} software~\cite{perseus}.

%\section*{TODO}
%\fxfatal{This list should be empty.}

\appendix
\section{Additional computational results} \label{sec:extra_experiments}
Some additional computational results have been relegated here for the
sake of readability.

\subsection{Comparison with multi-dimensional scaling} \label{sec:exp_poc_mds}
One may well ask whether the use of persistent homology actually
contributes anything useful to our proposed method. To investigate
this, we considered whether multi-dimensional scaling (MDS) could
successfully, \ie near-isometrically, embed the ``distances'' from the
various stages of the experiment in \secref{exp_poc} in
low-dimensional Euclidean space.

\begin{figure}[htbp]
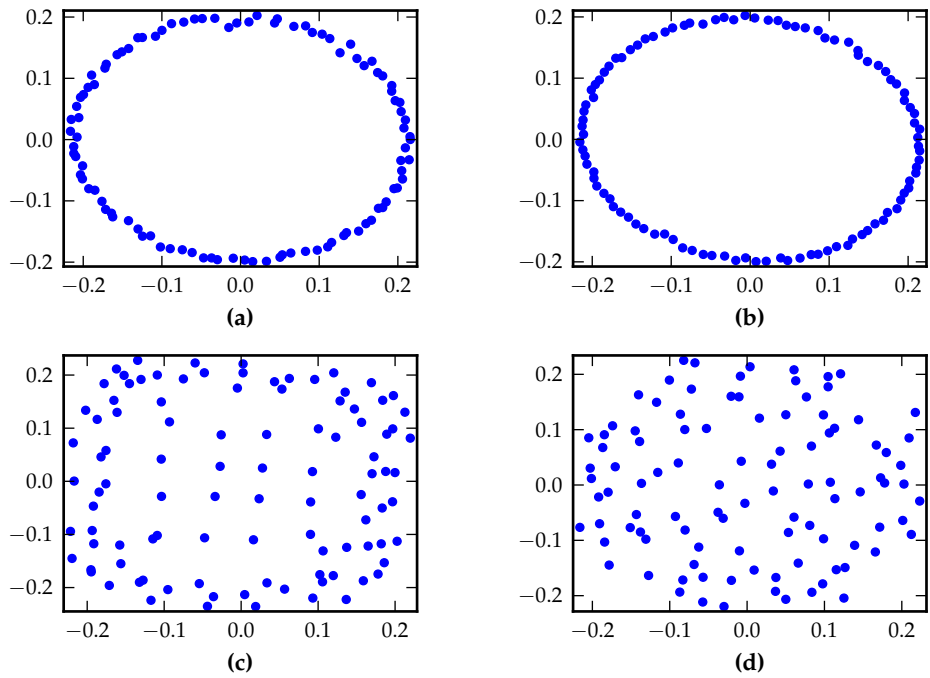

  \centering
  \begin{subfigure}[b]{0.45\textwidth}
    \fastfig{mds-poc}{cPsa_ordered_mds_60}
    \caption{} \label{subfig:exp_poc_mds_a}
  \end{subfigure}
  \begin{subfigure}[b]{0.45\textwidth}
    \fastfig{mds-poc}{cPpa-c,sp_ordered_mds_60}
    \caption{} \label{subfig:exp_poc_mds_b}
  \end{subfigure}
  \begin{subfigure}[b]{0.45\textwidth}
    \fastfig{mds-poc}{cPpa-c,h_ordered_mds_60}
    \caption{} \label{subfig:exp_poc_mds_c}
  \end{subfigure}
  \begin{subfigure}[b]{0.45\textwidth}
    \fastfig{mds-poc}{cPpa-c,sp,h_ordered_mds_60}
    \caption{} \label{subfig:exp_poc_mds_d}
  \end{subfigure}
  \caption{MDS embedding of the neuron distances from the experiment
    in \secref{exp_poc}.
    \mysubref{subfig:exp_poc_mds_a} Original data.
    \mysubref{subfig:exp_poc_mds_b} After removal of spatial tuning.
    \mysubref{subfig:exp_poc_mds_c} After removal of head direction tuning.
    \mysubref{subfig:exp_poc_mds_d} After removal of both covariates.
  } \label{fig:exp_poc_mds}
\end{figure}

\Figref{exp_poc_mds} shows that MDS provides the same useful
information as persistent homology before removal of covariates and
after removal of only spatial tuning. After removing head direction
tuning, however, MDS is unable to detect the annular nature of the
remaining spatial covariate (compare in particular
Figure~\ref{subfig:exp_poc_mds_c} and \figref{exp_poc_c,h_pd}). This
illustrates the essential role played by persistent homology in our
work.

We point out that embedding in $\mathbb{R}^3$ does not seem to
qualitatively improve the situation.

\newpage
\printbibliography

\end{document}